\documentclass[prb,twocolumn,floatfix,amsfonts]{revtex4-1}
\usepackage{graphicx,graphics,color,epsfig}% Include figure files
\usepackage{bm}
\usepackage{amsmath}
\usepackage{amssymb}
\usepackage{epstopdf}
\usepackage{epsfig}
\usepackage{graphicx}

\usepackage{longtable}
\usepackage{rotating}
\usepackage{multirow}

\begin{document}
\title{	Design and manipulation of high-performance photovoltaic systems based on two-dimensional novel KAgSe/KAgX(X=S,Te) van der Waals heterojunctions }

\author{Keying Han,\textit{$^{a}$} Qiang Wang,$^{\ast}$\textit{$^{a}$} Yan Liang,$^{\ast}$\textit{$^{b}$} Defeng Guo,$^{\ast}$\textit{$^{a}$} and Bin Wang$^{\ast}$\textit{$^{c}$}}
\address{$^a$ ~State Key Laboratory of Metastable Materials Science and Technology and Key Laboratory for Microstructural Material Physics of Hebei Province, School of Science, Yanshan University, Qinhuangdao 066004, People's Republic of China. E-mail: qingwang@ysu.edu.cn; guodf@ysu.edu.cn}
\address{$^b$ ~College of Physics and Optoelectronic Engineering, Faculty of Information Science and Engineering, Ocean University of China, Songling Road 238, Qingdao 266100, People's Republic of China. E-mail: yliang.phy@ouc.edu.cn}
\address{$^c$ ~Shenzhen Key Laboratory of Advanced Thin Films and Applications, College of Physics and Optoelectronic Engineering, Shenzhen University, Shenzhen, 518060, People's Republic of China. E-mail: binwang@szu.edu.cn}

\begin{abstract}
The realization of high-performance two-dimensional (2D) solar photovoltaic systems are both fundamentally intriguing and practically appealing to meet the fast-growing energy requirements. Since the limited application of single 2D crystals in photovoltaic, here we propose a family of 2D KAgSe/KAgX(X=S,Te) van der Waals heterostructures (vdWHs), which are constructed by combining two different KAgX layers through interlayer vdW interaction. After a systematic study and further regulatory research of these vdWHs based on the first-principles, numerous fascinating characteristics and physical mechanisms are obtained. Firstly, favorable potential applications of these vdWHs in photovoltaics are confirmed in virtue of their desirable optoelectronic properties, such as the robust stabilitis, moderate direct band gaps, type-II band alignments together with superior carrier mobilities, visible optical absorptions, power conversion efficiencys (PCEs) and photocurrents in their based photovoltaic devices. More importantly, when under varying vertical electric field ($E_{z}$), a phase transition of band alignment from type-II to type-I of these vdWHs can be induced by the opposite band shifts between layers, which may enrich their applications in light-emitting diodes and lasers. Meanwhile, the PCE can be expanded up to $23\%$, and an obvious red-shift peak of the photocurrent in the visible light range are also obtained at different $E_{z}$. These fascinating tunable properties of KAgSe/KAgX vdWHs under varying $E_{z}$ not only promote the improvement of their photoelectric performances, but the underlying mechanisms can also be applied to next experimental design and practical application of other 2D photovoltaic systems. Especially for the red-shift peak of the photocurrent, which is rarely found but highly desirable in practical visible photoelectric conversion.  
\end{abstract}

\maketitle

\section{Introduction}
Recent growth in the area of 2D crystals offers renewed opportunities for efficient and ultrathin solar photovoltaic systems, which can effectively utilize the solar energy.\cite{al2021solar,parida2011review} However, exploring novel 2D solar photovoltaic systems with high performance is still in great demand to meet the fast-growing energy requirements.\cite{allouhi2022up,tawalbeh2021} According to the different generation process of electron-hole pairs, the solar photovoltaic systems can be divided into two categories. One is based on the traditional bulk inorganic semiconductors, including Silicon, GaAs, and CdTe et.al.,\cite{papevz2020,LYJMCA4,2016large} in which the photoexcited electrons and holes are generated without intermediate steps, thus lacking of long-range Coulomb interaction. Despite tremendous efforts, their widespread applications are still hindered by various significant challenges, such as the instability and the overly cumbersome production process of silicon crystal solar cells, the heavy metal elements contained in CdTe, and the undersized energy conversion efficiency of GaAs and so forth. Another is the excitonic solar cells (XSCs), which is based on the donor–acceptor composite networks, and the excitons (electron–hole pairs) are generated and dissociated simultaneously at the donor–acceptor interface upon illumination.\cite{LYJMCA5,LYJMCA6} As indicated in previous studies, such excitons can be efficiently dissociated at the interfaces of XSC materials with different electron ionization and affinity potentials, thus substantial power conversion efficiencies (PCE) can be induced.\cite{ACS3} Even though, The PCE of most recent XSCs are still below 12\%, this is due to the fact that except for the promoting separation and hindering recombination of excitons, many other critical factors including effective visible light optical absorption, high carrier mobilities, small exciton binding energy, and moderate direct band gap ($1.2-1.6 eV$) of the donor, are also non-negligible.\cite{xu2020two,2018kagse} To this end, enormous challenges should still be overcome to design higher-performance XSCs, and searching new appropriate systems and exploring tunable mechanisms for improving the above critical factors still remain imperative.

Since the expeinmental mechanical exfoliation of graphene in 2004,\cite{gra2004} large amount of 2D materials have been reported in theory or laboratory one after another.\cite{2016h-BN,2018h-BN,2019MXenes,2017TMDS,2018TMDS} Due to the quantum confinement effect, 2D materials can exhibit more exotic properties and promising applications than their bulk counterparts,\cite{mak2010,20162d,gao20202d,zhang2020} making them more attractive for designing novel electronic and optoelectronic devices. Particularly, the recently reported new family of 2D ternary quintuple layers KAgX (X = S, Se, Te) have drawn tremendous interests for its thermoelectric and photovoltaic properties.\cite{mahmoud2019} The bulk KAgX has a PbFCl-type tetragonal structure with a space group of P4/nmm, and has been synthesized and characterized experimentally a decade ago.\cite{1981beitrage} Relevant researches and high-throughput calculations have demonstrated that the KAgSe monolayer can be easily exfoliated from its bulk phase in the experiment.\cite{2018kagse, xu2019new} Furthermore, a series of prominent photovoltaic behaviors have been verified in monolayer KAgX, including the moderate direct band gaps, high carrier mobilities, ideal visible light optical absorption coefficients, and remarkable performances in photovoltaic nano-devices.\cite{mahmoud2019} Even so, the invariable propertie of single 2D materials is far from enough to meet the urgent requirements of modern energy crisis. One effective way is constructing these existing 2D materials into heterojunctions,\cite{gobbi20182d,WQJMCA,su2020} which can enrich the exceptional properties of these isolated components and break the limitations of their applications in many fields.\cite{yang2018,xu2020s} Especiaslly in 2D XSCs, due to the interlayers charge transfer and accumulation, enhanced photovoltaic characteristics of the heterojunctions can be obtained than those of their components,\cite{WQFEPV,WQJMCA,zhou2016} and these superior characteristics of heterojunctions are also effectively tunable upon external pressure, electromagnetic field and illumination.\cite{gajdovs2006,yang2016,zhou2016,WQJMCC} Since the staggered band offset tends to separate the electrons and holes into different directions, the bilayer heterojunctions with type-II band alignment are dominant in 2D photoelectric devices and XSCs.\cite{18type,yu2013highly,xie2016promising} However, the investigation of KAgX heterojunctions for photovoltaics has been scarecely reported. Inspired by the ingenious works above and outstanding photovoltaic performances of 2D monolayer KAgX, we wonder if these excellent characteristics can be further enhanced by assembling them into van der Waals heterostructures (vdWHs). Through a further investigation on the tunable properties of these vdWHs, the final aim is to expand the regulatory mechanisms to next experimental design of 2D high-performance solar photovoltaic systems.

Here, based on the first-principles, the KAgSe/KAgX(X=S,Te) vdWHs are constructed and studied systematicially. It is found that these vdWHs behave staggered type-II band alignment due to the charge redistribution induced by the unique lattice stacking. Moreover, other numerous ideal characteristics are also found including the robust stabilitis, moderate direct band gaps, high carrier mobilities, efficient visible optical absorptions, considerable PCE and superior photocurrents in the 2D nano-devices, rendering them promising candidates for optoelectronic and photovoltaic devices. In light of the tunable interlayer charge transfer upon different vertical electric field ($E_{z}$), opposite band shifts occurs between the donor and acceptor layers. As a result, a phase transition from type-II to type-I band alignment is induced in such vdWHs, which may expand their applications in light-emitting diodes and lasers. More importantly, the above excellent photovoltaic characteristics of vdWHs could also be improved upon $E_{z}$, where the higher PCE, greater photocurrent and a red-shift peak of photocurrent in the visible light range can be obtained. These enhanced performances can further enrich their applications in photovoltaics, and the underlying mechanisms can also provide a theoretical guidance for futher experimental design and practical application of 2D photovoltaic systems, ecpecially for 2D XSCs.  

\section{Model and Numerical Method}
In this study, the structural, electronic and photoelectric behaviors of KAgSe/KAgX(X=S,Te) vdWHs were investigated as the Vienna ab initio simulation package (VASP), which is based on the density functional theory (DFT).\cite{JMCA40} During the geometrical optimisation and electronic structure self-consistent calculations, the exchange correlation functional was described using the Perdew-Burke-Ernzerhof (PBE) form in the generalised gradient approximation (GGA).\cite{JMCA41} The ion-electron interaction was manipulated by the projected augmented wave (PAW) method,\cite{PAW} together with a plane wave cut-off energy of 450 eV. The convergence criterions were $0.01 eV/${\AA} in force and  $10^{-5} eV$ in energy, the \textit{k}-point mesh of $15\times15\times1$ is used to sample the Brillouin zone, and a vacuum region of 18{\AA} was set to avoid the interaction between repeated images. In order to correct the band gap underestimate of PBE, the hybrid density functional of Heyd-Schuseria-Ernzerhof (HSE06)\cite{JMCA44} was employed under the Hartree-Fock exchange energy of 25$\%$. In addition, the dynamic stability was confirmed by the phonon spectrum, which was calculated by using the Nanodcal code. The thermal stability was verified through performing the \textit{ab initio} molecular dynamics (MD) simulations within a $2\times2\times1$ supercell under 300K, the lasting time was 5 ps a time step of 1 fs.  
\begin{figure*}[!tb]
	\includegraphics[width=15.1cm]{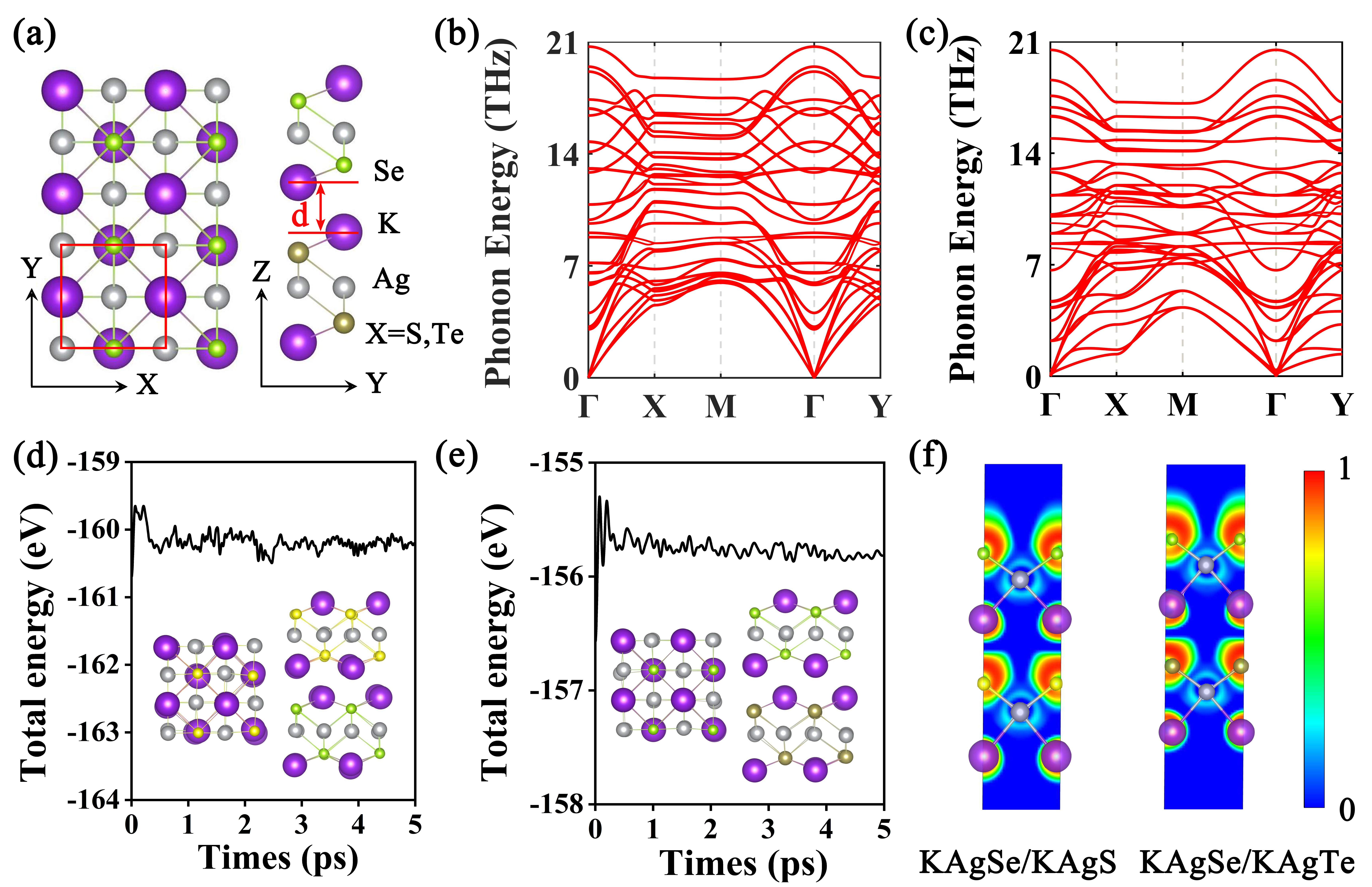}
	\centering
	\caption{ (a) Top and side views of the optimized atomic configurations of KAgSe/KAgX(X=S,Te) vdWHs, where the smallest unit cell is indicated by red square, and the equilibrium interlayer distance is labeled by d. (b-c) Phonon band dispersion curves of (b) KAgSe/KAgS and (c) KAgSe/KAgTe vdWHs. (d-e) The variation of total energies of the 2$\times$2 (d) KAgSe/KAgS and (e) KAgSe/KAgTe vdWHs supercells as a function of time at 300 K. Insets show the snapshot for each vdWH supercell at 5 ps from top and side views. (f) Side view of ElFs for each vdWH. }
	\label{fig1}
\end{figure*}
The photocurrents in KAgSe/KAgX(X=S,Te) based two-probe nanodevices were calculated by using the first-principles quantum transport package NanoDcal,\cite{nanodcal1,nanodcal2,KAS122} which is based on the combination of non-equilibrium Green's function and DFT (NEGF-DFT).\cite{nanodcal1} During the calculation, the norm-conserving nonlocal pseudo-potential was applied to define the atomic core.\cite{KAS51} The wave functions were expand by the basis sets with atomic orbitals of double-zeta polarization (DZP). The energy convergence criteria of self-consistence was set to be less than $10^{-5}$ eV. The generalized gradient approximation of PBE level was used to deal with the exchange-correlation potential.\cite{KAS52} In addition, a k-mesh gride of $24\times1$ was set perpendicular to the transport direction in the self-consistent calculation, and a $256\times1$ k-mesh gride was used during the computing of photocurrent.

\section{Numerical Results and Discussions}

\subsection{Structural characteristic and stability of KAgSe/KAgX (X = S, Te) vdWHs}

Monolayer KAgX (X=S, Se, Te) is a stratified material with alternating atomic layers of K-X-Ag-X-K along (001) direction.\cite{2018kagse,xu2019new} As is shown in Fig.~\ref{fig1} (a), different heterojunctions are constructed by stacking two adjacent KAgX monolayers with vdW interaction. Considerding the reasonable small lattice mismatch between each KAgX monolayers,\cite{xu2019new} 1$\times$1 unit cells of both patterns are chosen delicately to construct the vdWHs. The fully relaxed lattice constants are a=b=$4.398\AA$ for KAgSe/KAgS vdWH, corresponding to $\sim$3.24\% contraction of the pristine monolayer KAgSe and $\sim$1.17\% stretching of the pristine monolayer KAgS in the 2D plane; while for for KAgSe/KAgS vdWH, the fully optimized structural parameters are a=b=$4.637\AA$, corresponding to $\sim$4.03\% stretching and $\sim$1.65\% contraction of monolayer KAgSe and KAgTe, respectively. These scalings of lattice expansion and contraction are reasonable, which are less than the previously reported experimental feasibility of 4\% for the monolayer WS$_{2}$.\cite{2015strain} In addition, the optimized interfacial distances are $2.387\AA$ and $2.514\AA$ for KAgSe/KAgS and KAgSe/KAgTe vdWH, comparable to the equilibrium interlayer distance of bilayer KAgSe (2.483\AA), indicating a typical vdW interaction within these two heterojunctions. 

\begin{figure*}[!tb]
	\includegraphics[width=15.5cm]{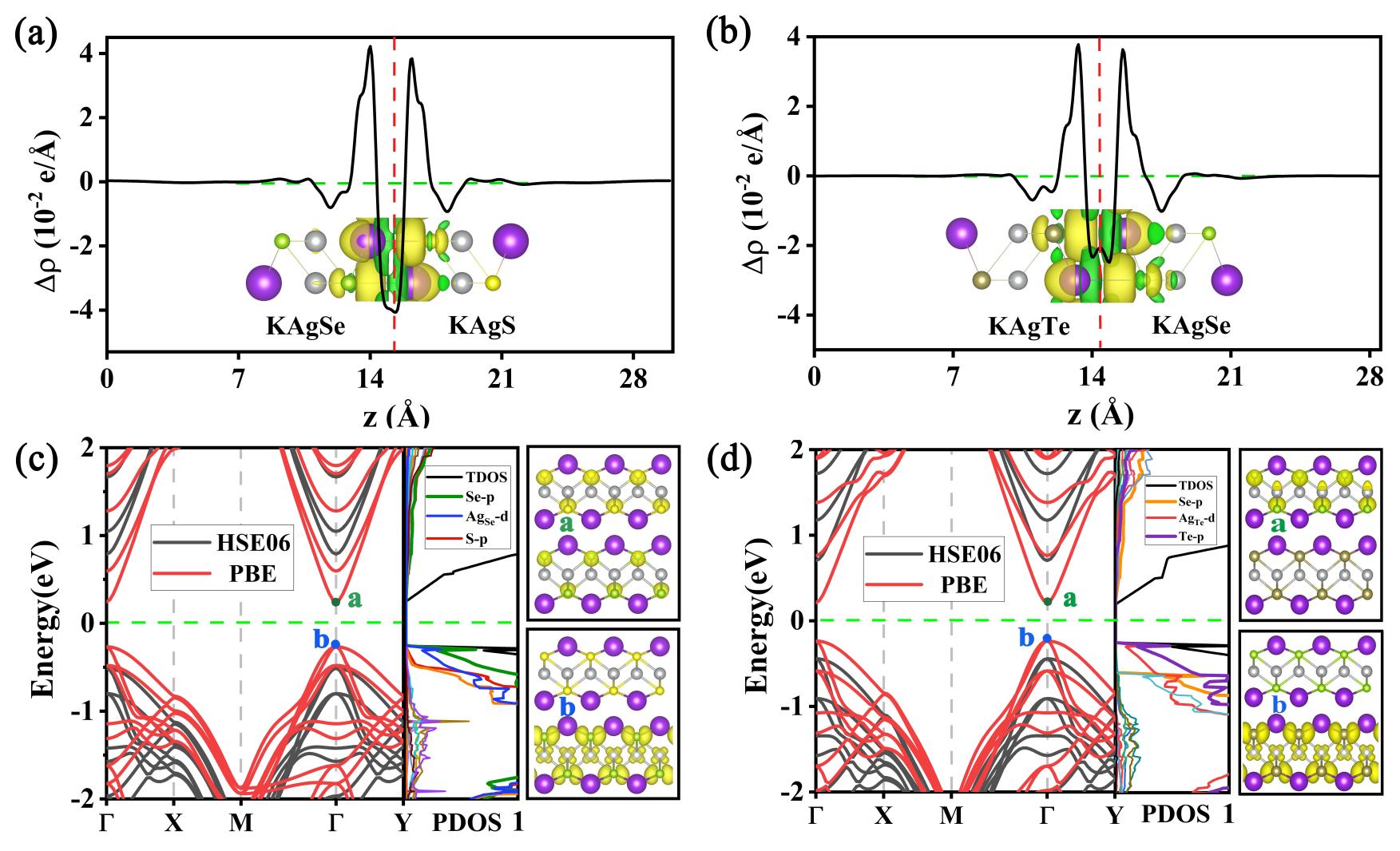}
	\centering
	\caption{ (a-b) Charge density deformations of plane projection and three-dimensional integration for (a) KAgSe/KAgS and (b) KAgSe/KAgTe vdWHs. The isosurface value is set to be 0.001 e{\AA}$^{-3}$, the fermi levels and interlayer interfaces are marked by horizontal green and vertical red dashed lines, respectively. (c-d) The calculated band structures (left panels), PDOS (middle panels), and partial charge densities at the CBM and VBM (right panels) of (c) KAgSe/KAgS and (d) KAgSe/KAgTe vdWHs, respectively. In each panel, the Fermi level of each system is set to zero and marked by the green horizontal dotted line. The band structures at PBE (HSE06) levels are labeled by red (black) lines, the DOS, PDOS, and partial charge densities are obtained at PBE level, and the value of the isosurface in each right panel is set at 0.005 e{\AA}$^{-3}$.}
	\label{fig2}
\end{figure*}

Before further evaluating the electric properties of these predictive KAgSe/KAgX vdWHs, their stabilities are checked carefully to explore the feasibility of their experimental synthesis. Firstly, the interfacial binding energy is calculated according to the folowing defination $E_{b}=(E_{vdw}-E_{top}-E_{bottom})/S$, where the  $E_{vdw}$, $E_{top}$ and $E_{bottom}$ are energies of the unit cell of the KAgSe/KAgX vdWH, the free-standing top and bottom monolayer, $S$ is the interface area. The calculated binding energy (Eb) are $-0.584 J/m^{2}$ for KAgSe/KAgS and $-0.538 J/m^{2}$ for KAgSe/KAgTe vdWH, larger than those of the recently reported WTe$_{2}$/HfS$_{2}$ ($-0.204 J/m^{2}$)\cite{jpcc121}, phosphorene/SnSe$_{2}$ ($-0.019 J/m^{2}$)\cite{PRA}, and KAgSe/SiC$_{2}$ ($-0.327 J/m^{2}$)\cite{WQJMCC}, indicating a relatively stronger interlayer interaction in these predictive KAgSe/KAgX vdWHs. Besides, the dynamical stability of these vdWHs is verified by calculating their phonon spectrums. As shown in Fig.~\ref{fig1} (b) and (c), the phonon spectrum contains no imaginary frequencies in the entire Brillouin zone, suggesting that the structures are stable minima on the potential energy surface, thus both KAgSe/KAgX vdWHs are dynamically stable.\cite{kagse50} Finally, the temporal evolution of the total energy is also calculated to inspect the thermal stabilities of these two vdWHs by using the first-principles MD simulation. For each KAgSe/KAgX vdWH, a 2 $\times$ 2 supercell is adopted for the simulation at 300 K, the evolutionary time reaches 5 ps with a time step of 1 fs. As shown in Fig.~\ref{fig1} (d) and (e), the total energies of KAgSe/KAgS and KAgSe/KAgTe vdWHs fluctuates around a constant value with considerably small fluctuation magnitude. The snapshots of the fully relaxed structure after 5 ps are also inserted in the corresponding figures, where no bond breaking or obvious geometry reconstruction occur, indicating that these two vdWHs are thermally stable at room temperature.

%\begin{figure*}[!tb]
%\includegraphics[width=16cm]{FigS1.jpg}
%\centering
%\caption{ (color online) Charge density deformation of plane projection and three-dimensional integration for the 2D bilayer 3R (a) BP, (b) BAs, and (c) BSb bilayers, respectively. In each panel, the red horizontal dotted lines denoted the Fermi level, the green vertical lines indicate edges of the top and bottom layers, and the yellow and cyan isosurfaces represent the charge accumulation and depletion at the interface with an isovalue of 0.015 e{\AA}$^{-3}$ (a, b) and 0.1 e{\AA}$^{-3}$ (c), respectively.}
%\label{figS1}
%\end{figure*}

%\begin{figure*}[!tb]
%\includegraphics[width=16cm]{FigS4.jpg}
%\centering
%\caption{ (color online) Plane averaged electrostatic potential of bilayer 2D 3R (a) BP, (b) BAs, and (c) BSb along the z direction. In each panel, the potential differences between top and bottom layers of the corresponding crystals are marked by $\Delta V$.}
%\label{figS4}
%\end{figure*}
\begin{figure*}[!tb]
	\includegraphics[width=15.2cm]{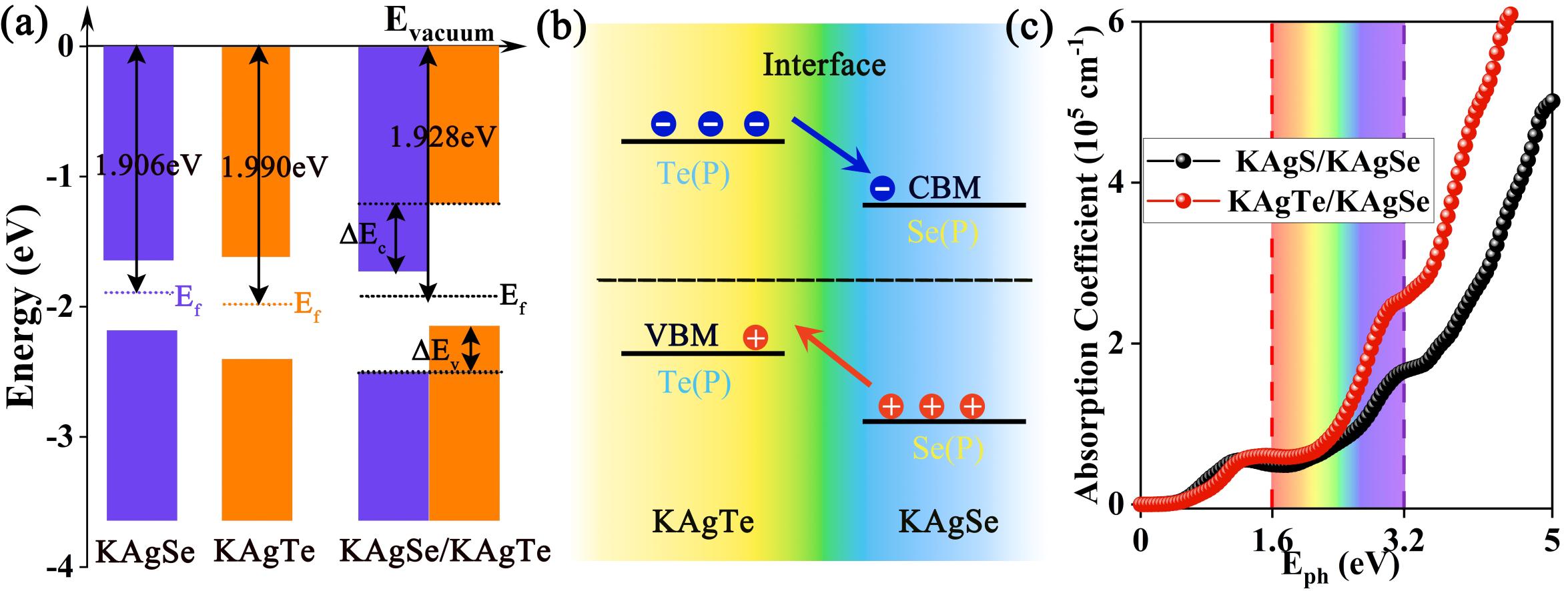}
	\centering
	\caption{(a) Schematic diagrams of the band alignment between KAgSe and KAgTe monolayers before and after forming the heterostructure. The work functions for each constructions are labelled, the conduction and valence band offset are also denoted by $\Delta E_{c}$ and $\Delta E_{v}$, respectively. (b) Schematic diagrams of the energy levels at the $\Gamma$ point of KAgSe/KAgTe vdWH, where the Fermi level is marked by black horizontal dotted line. (c) Optical absorption coefficients versus energy of the incident light for KAgSe/KAgX(X=S,Te) vdWHs.} 
	\label{fig3}
\end{figure*}
\subsection{Electronic properties of KAgSe/KAgX(X=S,Te) vdWHs}

In the following, we move to examine the electronic structures of the KAgSe/KAgX vdWHs. The bonding types of the KAgSe/KAgX vdWHs can be characterized by calculating the real space projection of electron localized functions (ELFs). As shown in Fig.~\ref{fig1} (f), the magnitude of ELF continuously varies from 0 to 1 describes the spatial localization of the reference electrons,\cite{JMCC_SiC2_55} where ELF = 1 or 0 represents complete localization or full decentralization of the electrons. For each monolayer component of the KAgSe/KAgX vdWHs, the ELFs between each pair of neighboring atoms are almost equal to zero, indicating an ionic interaction between them. Moreover, the weak vdW interfacial interaction can also be visualized by analyzing the corresponding ELF. The absence of electron localization suggests no atomiclly bonding at the interlayer region, no binds at the interface is formed and confirming a weak vdW interaction of the KAgSe/KAgX heterojunctions.

In order to gain insight into the interlayer charge transfer of KAgSe/KAgX vdWHs, their charge density differences $\Delta\rho(\textbf{r})$ are calculated based on the following equation,
\begin{equation}
	\Delta\rho(\textbf{r}) =[\rho_{KAgSe/KAgX}-\rho_{KAgSe}-\rho_{KAgX}]/n    
\end{equation}
where $\rho_{KAgSe/KAgX}$, $\rho_{KAgSe}$ and $\rho_{KAgX}$ are charge densities of KAgSe/KAgX vdWHs, KAgSe monolayer and KAgX monolayer, respectively, and n = 2 indicates the two different components of vdWHs. The 1D plane-averaged $\Delta\rho(\textbf{z})$ combine with the corresponding three-dimensional isosurface are shown in Fig.~\ref{fig2} (a) and (b). Obviously, distinct charge redistribution occurs at the interface. For KAgSe/KAgS vdWH, the charge mainly transfers from KAgS to KAgSe, while for KAgSe/KAgTe vdWH, the charge mainly transfers from KAgSe to KAgTe. this is resonable because of the increased work functions in the order of KAgS $<$ KAgSe $<$ KAgTe.\cite{WQJMCC} Detailed calculation of Bader charge indicates that roughly 0.0048 and 0.0066 electrons are transfered across the interface per unit cell of KAgSe/KAgS and KAgSe/KAgTe vdWH, respectively. As a result, the built-in electric field is larger at the interfacial region of KAgSe/KAgTe vdWH than that of the KAgSe/KAgS one.  

\begin{figure*}[!tb]
	\includegraphics[width=13.8cm]{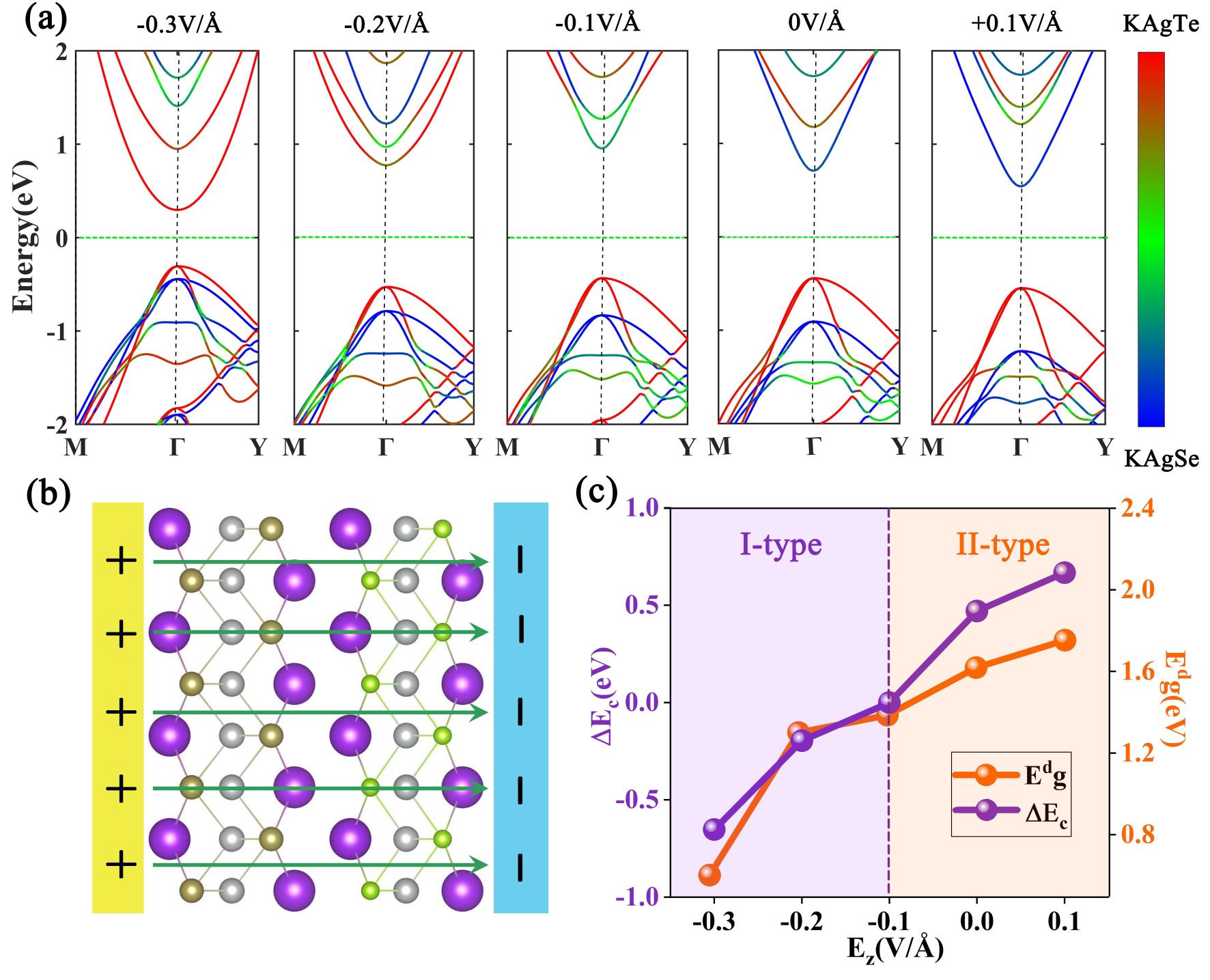}
	\centering
	\caption{(a) Projected band structures of KAgSe/KAgTe vdWH under different external electric fields at HSE06 level. In each panel, the contribution to the bands is projected to KAgSe and KAgTe indicated by different colors, and the green horizontal lines indicate the Fermi levels. (b) Schematic diagram of the KAgSe/KAgTe vdWH under a positive electric field. (c) Variation of the conduction band offest $\Delta E_c$ and donor band gap $E^{d}_g$ with external electric field at the HSE06 level. }
	\label{fig4}
\end{figure*}

After the detailed analysis of the electronic structures, we now focus on the band structures and density of states of the KAgSe/KAgX vdWHs. As shown in Fig. \ref{fig2} (c) and (d), the band structures, projected density of states (PDOS) and partial charge densities are depicted at PBE and HSE06 levels. Obviously, the band structures based on both functionals are qualitatively consistent, except for $\sim$ 0.7 eV underestimation of the bandgap at PBE level. Both KAgSe/KAgX vdWHs possess direct band gap semiconducting behaviors, with the conduction band minimum (CBM) and the valence band maximum (VBM) locating at $\Gamma$ point, as marked by "a" and "b" in the figure. Moreover, based on the HSE06 functional, the band gaps are 1.3 eV and 1.15 eV, respectively, for KAgSe/KAgS and KAgSe/KAgTe vdWHs, which almost perfectly satisfy the optimum range of 1.2$\sim$1.6 eV for excitonic solar cells.\cite{liang2018} These moderate ditect band gaps suggest KAgSe/KAgX vdWHs for potential applications in next 2D Solar photovoltaic materials. To understand the electronic band structures more clearly, the PDOS and the partial charge densities are presented in the middle and right panels in Fig.~\ref{fig2} (c) and (d). For KAgSe/KAgS vdWH, the CBM is dominated by S atoms of the top layer and Se atoms of bottom layer, while the VBM is mainly contributed by Se-$p$ and Ag-$d$ orbitals of the bottom layer. But for KAgSe/KAgTe vdWH, the CBM and VBM are almost solely dominated by the top and bottom layer, respectively. As a result, an obvious type-II band alignment is formed between the acceptor and donor owing to the spatial separation between CBM and VBM. This typical type-II band alignment can be seen more intuitively in the fourth panel of Fig.~\ref{fig4} (a), where the bands of different components are distinguished by different colors. 
To explore the physical mechanisms of the type-II band alignment of KAgSe/KAgTe vdWH, the schematic diagrams of the band alignment between KAgSe and KAgTe monolayers before and after forming the heterostructure are presented in Fig.~\ref{fig3} (a). the work functions for the KAgSe monolayer, KAgTe monolayer and KAgSe/KAgTe vdWH are 1.906 eV, 1.990 eV and 1.928 eV, respectively. As the work function of monolayer KAgSe is lower than that of KAgTe, when the KAgSe/KAgTe vdWH is formed, the charge tends to immigrate from KAgSe to KAgTe, which finally induce the built-in electric field at the interfaces. As a result, the work function is located exactly between the VBM of KAgTe and the CBM of KAgSe, forming a vdWH with staggered type-II band alignment. Futhermore, both the CBM and VBM of KAgSe layer are lower than those of the KAgTe layer, and the CBM offset $\Delta E_{c}$ and VBM offset $\Delta E_{v}$ are calculated to be 0.47 eV and 0.46 eV, respectively.

Especially, this kind of staggered type-II band alignment is confirmed host great adavantagies for the separation of excitons, thus is appealing in the XSCs\cite{JMCA53}. For KAgSe/KAgTe vdWH as demonstrated in Fig.~\ref{fig3} (b), the free electrons and holes tend to separate into different layers. when a beam of light with an appropriate frequency is irradiated, the photo-induced electrons transfer from the KAgTe layer to the KAgSe, accompanied with the holes moving towards the opposite direction. This reduces the possibility of electron–hole recombination, and eventually promoting the efficiency of electron‐hole pairs separation in the real space. Moreover, a new built-in electric field is formed during the charge redistribution in light of the domain separation of electrons and holes, which also facilitates the separation of the photo-excited electron–hole pairs by weakening the Coulomb interaction between them. All these promising features endows type-II KAgSe/KAgTe vdWH appealing applications in 2D photovoltaic devices and XSCs.
\begin{figure*}[tb]
	\includegraphics[width=14.6cm]{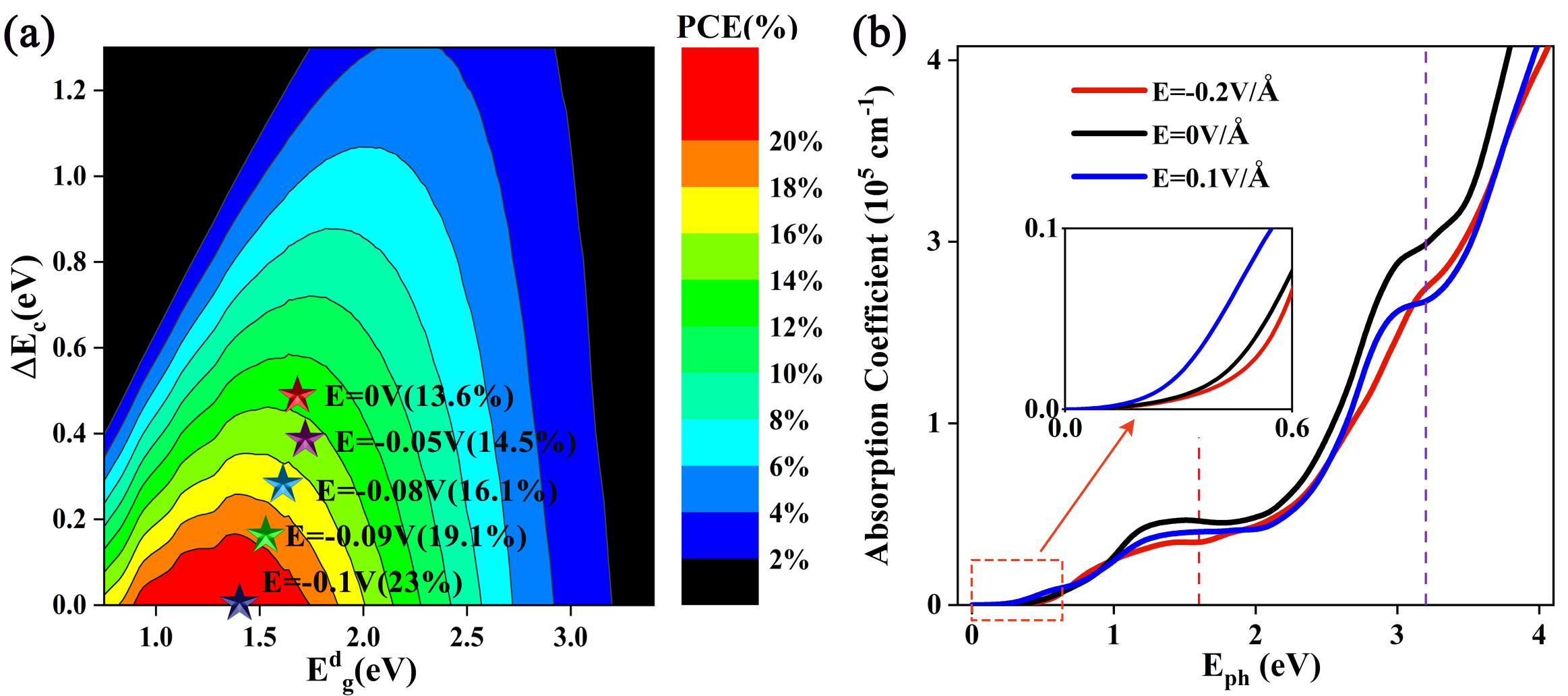}
	\centering
	\caption{(a) Contour plot of power conversion efficiency (PCE) as a function of conduction band offest $\Delta E_c$ and donor band gap $E^{d}_g$ under different external electric fields. (b) Variation of optical absorption coefficient with respect to the external electric field, and inset shows the curve of the absorption of incident light energy around the bandgap. }
	\label{fig5}
\end{figure*}
\subsection{optical absorptions and Carrier mobilities of KAgSe/KAgX(X=S,Te) vdWHs}

In light of the potential applications of KAgSe/KAgX vdWHs in 2D photovoltaics, now we turn to evaluate their optical harvesting abilities by calculating the optical absorption coefficient $a(\omega)$, which can be expressed as follows,\cite{JMCA560}
\begin{equation}
	a(\omega) = \frac{\sqrt{2}\omega}{c} \sqrt{\sqrt{\epsilon_1^2(\omega)+\epsilon_2^2(\omega)}-\epsilon_1(\omega)},
\end{equation}
where $\omega$ denotes the frequency of incident light, $c$ is the velocity of light, $\epsilon_1$ and $\epsilon_2$ are real and imaginary parts of the dielectric function, respectively. During the calculation, $\epsilon_2$ is calculated using a summation over empty states, while $\epsilon_1$ can be obtained through the Kramers-Kronig transformation.\cite{gajdovs2006} As shown in Fig.~\ref{fig3} (c), the absorption coefficients of both KAgSe/KAgX vdWHs start to vibrate when the light energy increases above the value of band gaps, and then possesses high absorption over a wide range of visible light ($\sim$1.6-3.2 eV). The maximum values can reach $1.65 \times10^{5} cm^{-1}$ of KAgSe/KAgS and $2.56 \times10^{5} cm^{-1}$ of KAgSe/KAgTe vdWHs, nearly one order of magnitude larger than those of the ideal 2D layered MoSSe\cite{lyJMCA} and InSe\cite{Jinhaopccp}, indicating preferable solar harvesting abilities of the KAgSe/KAgX vdWHs.  
\begin{table}[ht]
	\centering
	\caption{ carrier mobilities ($\mu _{2D}$) of KAgSe/KAgX vdWHs for electron, light hole, and heavy hole.}
	\begin{tabular}{lccccccccc}
		\hline
		   crystal structures          & $ e $       & $h_{light}  $          & $ h_{heavy} $      \\
		\hline
		$KAgSe/KAgS  (cm^{2}V^{-1}s^{-1})$  &  2475.105  & 2031.232   &   13.205    \\
		$KAgSe/KAgTe (cm^{2}V^{-1}s^{-1})$  &  2846.714   & 2055.625   &   11.813    \\
		\hline
	\end{tabular}%
	\label{tab:table1}%
\end{table}%

Since the carrier mobility ($\mu_{2D}$) is another crucial factor affecting the efficiency of photovoltaic devices, we have also predicted the $\mu_{2D}$ of KAgSe/KAgX vdWHs based on the deformation potential (DP) theory,\cite{JMCA631} and the theoretical reliability of DP theory has been widely examined in many 2D atomic structures.\cite{2006design,li2023type}. 
\begin{equation}
	\mu_{2D}=\frac{e\hbar^{3}C_{2D}}{k_{B}T|m_{e/h}^{*}|^{2}E_{1}^{2}}.
	\label{cur1}
\end{equation}
here e is the electron charge, $C_{2D}$ is the elastic modulus, $k_{B}$ is the Boltzmann constant, $T$ is the temperature, $m_{e/h}^{*}$ is the effective mass of electrons or holes, and $E_{1}$ is the deformation potential. The obtained carrier mobilities of KAgSe/KAgX vdWHs are displayed in Table~\ref{tab:table1}, which are approximately consistent with those of monolayer KAgS ($6.257\times 10^{3}cm^{2}\cdot V^{-1}\cdot s^{-1}$)\cite{2018kagse} and KAgSe ($4.43\times 10^{3}cm^{2}\cdot V^{-1}\cdot s^{-1}$) \cite{xu2019new}. Although lower than the previous first principles results of graphene ($3.57\times 10^{5}cm^{2}\cdot V^{-1}\cdot s^{-1}$), but much larger than those of 2D Si$(\sim480-1300\times 10^{2} cm^{2}\cdot V^{-1}\cdot s^{-1})$\cite{shao2013}, MoS$_{2}$($\sim200-500\times 10^{2} cm^{2}\cdot V^{-1}\cdot s^{-1}$)\cite{2011single}, and chalcogenides$(\sim300-2000\times 10^{2} cm^{2}\cdot V^{-1}\cdot s^{-1})$\cite{jin2016}. These high carrier mobilities can further highlight their potential application prospects in electronics and photoelectronics.
 \begin{figure*}[tb]
	\includegraphics[width=15cm]{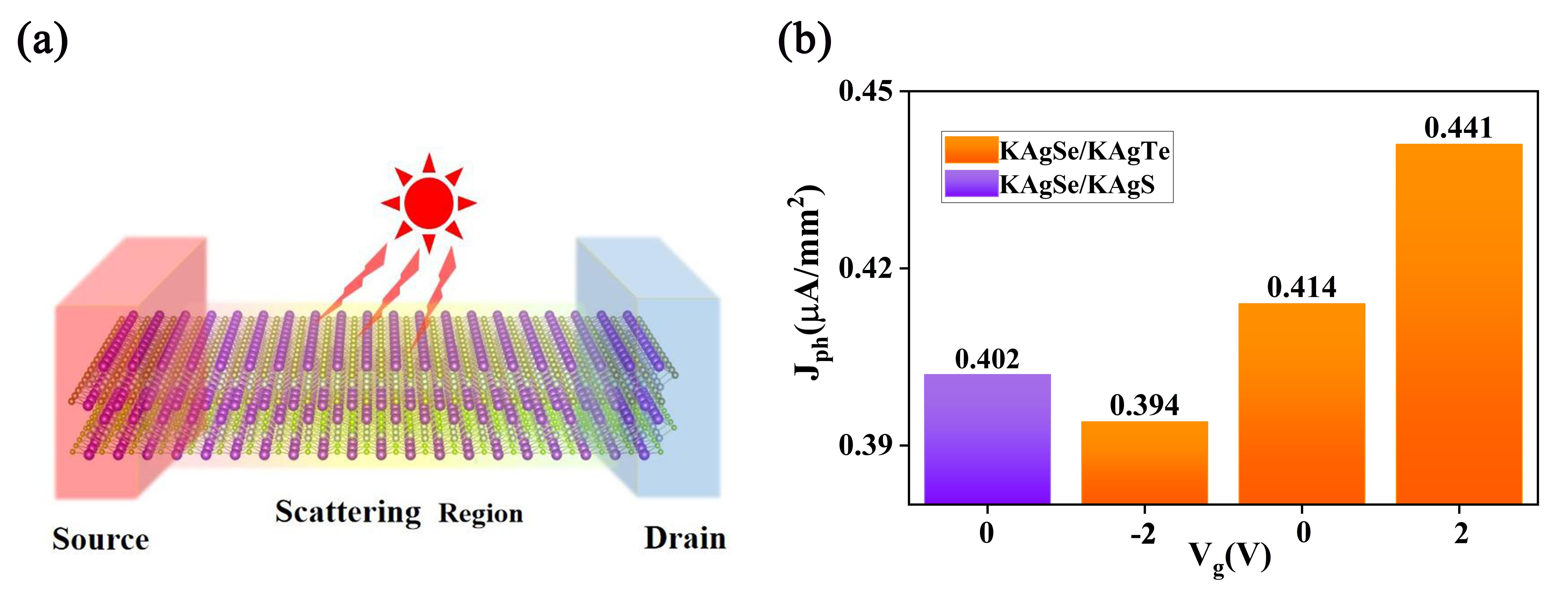}
	\centering
	\caption{(a) Schematic of the 2D KAgSe/KAgX vdWHs based two-probe photovoltaic device. (b) Photocurrents of KAgSe/KAgX vdWHs based nano devices under different gate voltage ($V_{g}$), with the incident photon energy around the corresponding band gap.}
	\label{fig6}
\end{figure*}
\subsection{ Tunable effect of external electric field on KAgSe/KAgTe vdWH }

As mentioned above, the electronic and photovoltaic behaviors of the KAgSe/KAgX vdWHs mianly depend on the charge transfer across the interface, which can be potentially tuneable under various external conditions. In experiments, the vertica l external electric field ($E_{Z}$) is an effective method of manipulation.\cite{InSePRB,chen2019tunable} We thus herein investigate the KAgSe/KAgTe vdWH as a representative to regulate its photovoltaic performance under applied vertical external electric field. As shown in Fig.~\ref{fig4} (b), an external electric field is applied perpendicular to the 2D plane along the direction from KAgTe to KAgSe, which is in the same direction as that of the built-in electric field. The projected band structures of KAgSe/KAgTe vdWH under different external electric fields are shown in Fig.~\ref{fig4} (a), where the blue and red curves indicate the bands contributed by KAgSe and KAgTe layer, respectively. Obviously, both CBM and VBM of KAgSe layer shift upward as $E_{Z}$ decreases from $0.1 V/${\AA} to $-0.3 V/${\AA}, accompany with an obvious downward move appears for the CBM of the KAgTe layer. This is reasonable because of the opposite directions of external and built-in electric fields, where the negative increasing external electric field will hinder the relative movement of band edges between KAgSe and KAgTe layers caused by the built-in field as has discussed in Fig.~\ref{fig4} (c). More interestingly, when $E_{Z}$ decreases smaller than $-0.1 V/${\AA}, the CBM of KAgSe decreases across that of KAgTe, suggesting a transition of band alignments from type-II (staggered gap) to type-I (straddling gap). For type-I vdWHs, electrons and holes are more preferentially transfered in the same layer of the vdWH, which will promote the recombination of charge carriers and reduces their separation. Therefore, the type-I vdWHs are no longer conducive to applications in solar energy conversion, but are suitable for luminescent devices, including the light-emitting lasers and diodes.\cite{1995high,1986quantum}

Next, we explore the tunable photovoltaic performances of the KAgSe/KAgTe vdWH under different $E_{Z}$. PCE is a key factor to describe the efficiency of XSC by quantifying the percentage of solar energy converted into power, which can be calculated by \cite{2006design},
\begin{equation}
	\begin{aligned}
		PCE =\frac{0.65(E^{d}_{g}-\Delta E_{c}-0.3)\int_{E^{d}_{g}}^{\infty}\frac{P(\hbar \omega)}{\hbar \omega}d(\hbar \omega)}{\int_{0}^{\infty}P(\hbar \omega)d(\hbar \omega)},
	\end{aligned}  \label{eq1}
\end{equation}
Here the band-fill factor (FF) is assumed to be 0.65.\cite{2006design} $E^{d}_g$ is band gap of the donor segment, and $\Delta E_c$ is the magnitude of conduction band offset, which can be obtained by the difference of CBM between the donor and acceptor segments. $(E^{d}_{g}-\Delta E_{c}-0.3)$ indicates the estimated value of the maximum open circuit voltage; and $P(\hbar \omega)$ is the solar energy flux of $AM1.5$ at the photon energy of $\hbar \omega$. The integral formulas at the numerator and the denominator are the short-circuit current and the total incident solar power per unit area, respectively. As shown in Fig.~\ref{fig4} (c), $E^{d}_g$ and $\Delta E_c$ are also tunable by varing $E_{Z}$. When $E_{Z}$ is decreased from $0.1 V/${\AA} to $-0.1 V/${\AA}, the type-II band alignment of KAgSe/KAgTe vdWH is well maintained, and a significant increase of PCE can be obtained by the rapid reduction of $\Delta E_c$ and $E^{d}_g$. Fig.~\ref{fig5} (a) shows the PCE phase diagram of KAgSe/KAgTe vdWH versus $E^{d}_g$ and $\Delta E_c$ under different $E_{Z}$. At the equilibrium external electric field ($E_{Z}$=0), The PCE can be reached 13.6\% with $\Delta E_c$=0.47 eV and $E^{d}_g$=1.62 eV, which is greater than that of the best certified efficiency of the organic solar cell (11.7\%)\cite{JMCA61} and the typical MoS$_{2}$/p-Si heterojunction solar cell.\cite{JMCA62} More intriguingly, the PCE can be improved to 23.0\% when $E_{Z}$ is decreased to $-0.1 V/${\AA}, and such high PCE has outperformed than those of most available materials so far.\cite{zhoubinJMCC} In addition, considering the decisive impact of optical absorption coefficient ($a(\omega)$) on photovoltaics, the values of $a(\omega)$ under varying $E_{Z}$ is further studied. As is shown in Fig.~\ref{fig5} (b), the shape of $a(\omega)$ under different $E_{Z}$ are obtained similar to the pristine vdWH under $E_{Z}$=0. When $E_{Z}$ increases from $-0.2 V/${\AA} to $0.1 V/${\AA}, a blue-shift of $a(\omega)$ occurs under $E_{Z}<0$ and red-shift under $E_{Z}>0$ at the energy of incident light around the band gaps, this resulted from the broadening and narrowing band gaps under $E_{Z}<0$ and $E_{Z}>0$, respectively. However, within the visible light range, a fascinating blue-shift of $a(\omega)$ can be detected for both $E_{Z}<0$ and $E_{Z}>0$ due to the relative offset of energy bands between KAgSe and KAgTe layers. All these excellent tunable performances of solar absorption and PCE indicate a highly desirable XSCs application of KAgSe/KAgTe vdWH, which may also lead to the adjustability of photocurrent within the visible light range of next photovolatic devices.
\begin{figure*}[tb]
	\includegraphics[width=15cm]{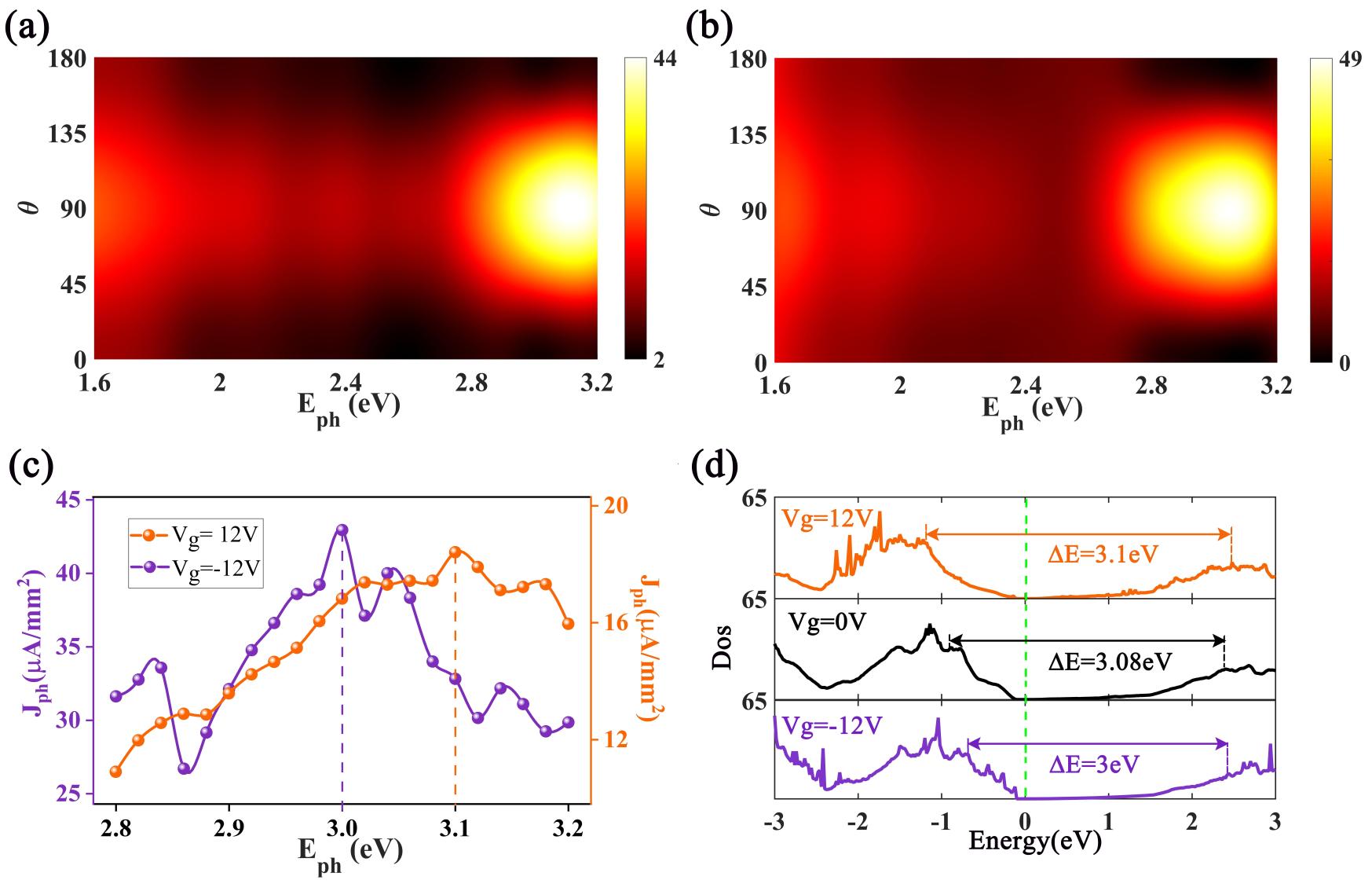}
	\centering
	\caption{The calculated densitys of photocurrent ($J_{ph}$) versus incident light energy ($E_{ph}$) and polarization angle $\theta$ for (a) KAgSe/KAgS and (b) KAgSe/KAgTe vdWHs, respectively. During our calculation, the power density ($P_{in}$) of the incident light is set 10$^{3}$ $\mu Wmm^{-2}$). (c) $J_{ph}$ of KAgSe/KAgTe vdWH versus $E_{ph}$ at $\theta=90^{\circ} $ under $V_{g}=\pm 12 V$. (d) Density of states (Dos) versus energy for KAgSe/KAgTe vdWH under different $V_{g}$, where the Dos under $V_{g}=-12, 0, 12 V$ are marked by orange, black, purple curves, respectively.}
	\label{fig7}
\end{figure*}

\subsection{Photocurrent in different types of KAgSe/KAgX(X=S,Te) vdWHs based nanodevices}
Although the excellent and tunable photovoltaic properties of KAgSe/KAgX vdWHs have been revealed, there are still qualitative predictions due to the limitation of periodic boundary condition. Since the quantum scattering has to be solved under the open boundary condition in practical photoelectric nanodevices, it is necessary to build the KAgSe/KAgX vdWHs based nanodevices to investage their actual photovoltaic performances. As is shown in Fig.~\ref{fig6} (a), the schematic structure of a KAgSe/KAgX vdWH based two-probe photovoltaic device is presented, where both leads are constructed by periodic extension of the scattering region, and the whole scattering region is irradiated by a vertical linearly polarized light. Such simplified device model has been employed extensively in experimental battery design, and posess numerous advantages than the traditional metal-simiconductor tunneling junctions, including the smooth and continuous interface,\cite{WQJMCC,PRA,WQJMCA,2018kagse,liang2018} simple structure, perfect lattice matching, easy device preparation and so forth. A small bias voltage between the source and drain is applied to drive the flowing of photocurrent, which is far less than the band gap. When the incident photon energy (E$_{ph}$) is greater than the band gap, electrons in the the valence band can be excited into the the conduction band accompany with the generation of electron-hole pairs. Then, the electron and hole separated along opposite directions under driven of the potential difference between source and drain, which finally induces the formation of photocurrent. At the first order of Born approximation, the photocurrent flowing into the left probe can be expressed as,\cite{JMCA65,JMCA66,JMCA67}
\begin{equation}
	\begin{aligned}
		I_{L}^{ph}=\frac{ie}{h}\int \textbf{Tr}\left[\Gamma_{L}\{G^{<(ph)}+f_{L}(E)(G^{>(ph)}-G^{<(ph)})\}\right]dE,
		%J_{L}^{ph}=&\frac{ie}{h}\int \textbf{Tr}\left[\Gamma_{L}\{G^{<(ph)}+f_{L}(E)(G^{>(ph)}-G^{<(ph)})\}\right]dE,
	\end{aligned}  \label{cur0}
\end{equation}
where $\Gamma_L$ is the line-width function, which represents the coupling between the left lead and the central scattering region; $G^{>/<(ph)}$ is the greater/lesser Green’s function considering the electron–photon interaction. $f_L$ is the Fermi distribution function of the left lead.\cite{JMCA66,JMCA67} 

As is shown in Fig.~\ref{fig6} (b), The calculated photocurrent density $J_{ph}=I_L^{ph}/S$ of KAgSe/KAgX vdWHs based nanodevices at the E$_{ph}$ equal to the corresponding band gaps are displayed. Due to the stronger built-in electric field and higher absorption coefficient, a higher $J_{ph}$ in the KAgTe/KAgSe based device can be detected than that of KAgSe/KAgS vdWH. Furthermore, considering the tunability of band alignment types and PCE under different vertical external electric field as discussed above, a gate voltage $V_{g}$ is also applied perpendicularly across the scattering region of KAgTe/KAgSe device to regulate its $J_{ph}$. As expected, $J_{ph}$ decreased obviously as the $V_{g}$ reduces from $2 V$ to $-2 V$. This resulted from the weakened built-in electric field and type-II to type-I band alignment changes under the negative $V_{g}$, which finally hindering the values of photocurrent. 

Next, we expand E$_{ph}$ to the visible light region and discuss the variation of relationship between $J_{ph}$ and polarization 
angle of E$_{ph}$. As displayed in Fig.~\ref{fig7} (a) and (b), broad peaks of $J_{ph}$ can be obtained in the visible light region. The maximum values can be reached 49 $\mu Amm^{-2}$ for KAgSe/KAgTe and 44 $\mu Amm^{-2}$ for KAgSe/KAgS devices, which are larger than those of MoS$_{2}$, and GeS devices under the same external conditions,\cite{zhao2017design} indicating appealing photoelectric conversion efficiencies of these vdWH devices in the visible light region. Furthermore, the higher $J_{ph}$ of KAgSe/KAgTe than that of KAgSe/KAgS device can be attributed to its stronger built-in electric field and more effective $a(\omega)$ in the visible light region. Moreover, the distribution of $J_{ph}$ exhibits symmetric relative to $\theta=90 ^{\circ}$ with the maximum values locating at $\theta=90 ^{\circ}$ for both devices. Since the expression of $J_{ph}$ can be linearly decomposed into three terms, which are proportional to $sin^{2}\theta$, $cos^{2}\theta$ and $sin2\theta$, respectively, and the distribution of $J_{ph}$ is finally determined by their competition.\cite{JMCA66} Finally, the influence of $V_{g}$ on the $J_{ph}$ of its device in the visible light region is further explored. As shown in Fig.~\ref{fig7} (c), when $V_{g}$ = 0 V, the maximum $J_{ph}$ of KAgSe/KAgTe device appears at E$_{ph}$= 3.08 eV, and an obvious red-shift and blue-shift of $J_{ph}$ peak can be observed under $V_{g}$ = -12V and 12V, respectively. Here, we should mention that due to the distinct electron-hole recombination rate and quantum transport behaviors in each 2D device,\cite{jinHao88} the different peak shifts between $J_{ph}$ and $a(\omega)$ in the visible light range is reasonable, and the actual energy difference between each $J_{ph}$ peak is also close to the energy difference between the two peaks of the corresponding density of states near the Fermi energy as presented in Fig.~\ref{fig7} (d). All these high performances and tunable characteristics suggest that 2D KAgSe/KAgX vdWHs possess powerful potential in 2D optoelectronic and photovoltaic applications, and the physical mechanisms of red-shift peak of $J_{ph}$ in the visible light region may shed a new light on next experimental design of 2D high performance photovoltaic devices.
 
%\begin{figure*}[!tb]
%\includegraphics[width=16cm]{FigS5.jpg}
%\centering
%\caption{(color online) The density of states (Dos) versus energy of the monolayer BP (red curve) and bilayer 3R BP (green curves) structures under d=4.1{\AA}, 3.5{\AA}, and 2.9{\AA}, respectively. The green dash lines in each panel indicate the fermi levels.}
%\label{figS5}
%\end{figure*}

%and $\eta$ is defined by the ratio of maximum output photoelectron power $P_{out}$ density to the incident one $P_{in}$:
%\begin{equation}
%\eta=P_{out}^{max}/P_{in}
%\end{equation}
%Here, $P_{out}$ of the devices can be obtained by the product of $V_{d}$ and the corresponding $J_{ph}$.

{\section{Discussion}}
In summary, we have studied the high performances and feasibility of photovoltaic based on the KAgSe/KAgX vdWHs. Their high mechanical, dynamic and thermal stability ensure high feasibilities of experimental synthesis. Their moderate direct band gaps, type-II band alignments, high carrier mobilities, efficient visible optical absorptions, ideal PCE and excellent photocurrent in the devices make them promising candidates for 2D high-performance photovoltaic devices. More interestingly, a phase transition of band alignment from type-II to type-I of the vdWH, combine with the further increased PCE (up to $23\%$) and a red-shift peak of the photocurrent in the visible light region are also detected under varying vertical electric field, which not noly significantly improve and enrich the photovoltaic performances of these KAgSe/KAgX vdWHs, but also shed light on the investigations on to other similar 2D photovoltaic heterojunctions. Therefore, these calculations illustrate new avenues for next experimental design and manipulation of novel 2D solar photovoltaic systems.
\section*{Conflicts of interest}

The authors declare no competing financial interest.

\section*{Acknowledgements}

This work was financially supported by grants from the National Natural Science Foundation of China (Grant No.11774238 and 12034014), Shenzhen Natural Science Foundations (Grant No. JCYJ20190808150409413), and Y.L acknowledges the Young Talents Project at Ocean University of China.
%\bigskip
%\noindent{$^{\dag)}$ binwang@szu.edu.cn} \\
%\bigskip

%%%END OF MAIN TEXT%%%


%\balance


%%%REFERENCES%%%
\bibliography{main} %You need to replace "rsc" on this line with the name of your .bib file

\providecommand*{\mcitethebibliography}{\thebibliography}
\csname @ifundefined\endcsname{endmcitethebibliography}
{\let\endmcitethebibliography\endthebibliography}{}
\begin{mcitethebibliography}{75}
\providecommand*{\natexlab}[1]{#1}
\providecommand*{\mciteSetBstSublistMode}[1]{}
\providecommand*{\mciteSetBstMaxWidthForm}[2]{}
\providecommand*{\mciteBstWouldAddEndPuncttrue}
  {\def\EndOfBibitem{\unskip.}}
\providecommand*{\mciteBstWouldAddEndPunctfalse}
  {\let\EndOfBibitem\relax}
\providecommand*{\mciteSetBstMidEndSepPunct}[3]{}
\providecommand*{\mciteSetBstSublistLabelBeginEnd}[3]{}
\providecommand*{\EndOfBibitem}{}
\mciteSetBstSublistMode{f}
\mciteSetBstMaxWidthForm{subitem}
{(\emph{\alph{mcitesubitemcount}})}
\mciteSetBstSublistLabelBeginEnd{\mcitemaxwidthsubitemform\space}
{\relax}{\relax}

\bibitem[Al-Shahri \emph{et~al.}(2021)Al-Shahri, Ismail, Hannan, Lipu,
  Al-Shetwi, Begum, Al-Muhsen, and Soujeri]{al2021solar}
O.~A. Al-Shahri, F.~B. Ismail, M.~Hannan, M.~H. Lipu, A.~Q. Al-Shetwi,
  R.~Begum, N.~F. Al-Muhsen and E.~Soujeri, \emph{J Clean Prod}, 2021,
  \textbf{284}, 125465\relax
\mciteBstWouldAddEndPuncttrue
\mciteSetBstMidEndSepPunct{\mcitedefaultmidpunct}
{\mcitedefaultendpunct}{\mcitedefaultseppunct}\relax
\EndOfBibitem
\bibitem[Parida \emph{et~al.}(2011)Parida, Iniyan, and Goic]{parida2011review}
B.~Parida, S.~Iniyan and R.~Goic, \emph{Renew. Sust. Energ. Rev}, 2011,
  \textbf{15}, 1625--1636\relax
\mciteBstWouldAddEndPuncttrue
\mciteSetBstMidEndSepPunct{\mcitedefaultmidpunct}
{\mcitedefaultendpunct}{\mcitedefaultseppunct}\relax
\EndOfBibitem
\bibitem[Allouhi \emph{et~al.}(2022)Allouhi, Rehman, Buker, and
  Said]{allouhi2022up}
A.~Allouhi, S.~Rehman, M.~S. Buker and Z.~Said, \emph{J Clean Prod}, 2022,
  132339\relax
\mciteBstWouldAddEndPuncttrue
\mciteSetBstMidEndSepPunct{\mcitedefaultmidpunct}
{\mcitedefaultendpunct}{\mcitedefaultseppunct}\relax
\EndOfBibitem
\bibitem[Tawalbeh \emph{et~al.}(2021)Tawalbeh, Al-Othman, Kafiah, Abdelsalam,
  Almomani, and Alkasrawi]{tawalbeh2021}
M.~Tawalbeh, A.~Al-Othman, F.~Kafiah, E.~Abdelsalam, F.~Almomani and
  M.~Alkasrawi, \emph{Sci Total Environ}, 2021, \textbf{759}, 143528\relax
\mciteBstWouldAddEndPuncttrue
\mciteSetBstMidEndSepPunct{\mcitedefaultmidpunct}
{\mcitedefaultendpunct}{\mcitedefaultseppunct}\relax
\EndOfBibitem
\bibitem[Pape{\v{z}} \emph{et~al.}(2020)Pape{\v{z}}, Gajdo{\v{s}}, Dallaev,
  Sobola, Sedlak, Mot{\'u}z, Nebojsa, and Grmela]{papevz2020}
N.~Pape{\v{z}}, A.~Gajdo{\v{s}}, R.~Dallaev, D.~Sobola, P.~Sedlak,
  R.~Mot{\'u}z, A.~Nebojsa and L.~Grmela, \emph{Appl Surf Sci}, 2020,
  \textbf{510}, 145329\relax
\mciteBstWouldAddEndPuncttrue
\mciteSetBstMidEndSepPunct{\mcitedefaultmidpunct}
{\mcitedefaultendpunct}{\mcitedefaultseppunct}\relax
\EndOfBibitem
\bibitem[Ginley \emph{et~al.}(2008)Ginley, Green, and Collins]{LYJMCA4}
D.~Ginley, M.~A. Green and R.~Collins, \emph{Mrs Bull}, 2008, \textbf{33},
  355--364\relax
\mciteBstWouldAddEndPuncttrue
\mciteSetBstMidEndSepPunct{\mcitedefaultmidpunct}
{\mcitedefaultendpunct}{\mcitedefaultseppunct}\relax
\EndOfBibitem
\bibitem[Luo \emph{et~al.}(2016)Luo, Liu, Yang, Bao, Li, Zhang, Li, Wu, and
  Feng]{2016large}
R.~Luo, B.~Liu, X.~Yang, Z.~Bao, B.~Li, J.~Zhang, W.~Li, L.~Wu and L.~Feng,
  \emph{Appl Surf Sci}, 2016, \textbf{360}, 744--748\relax
\mciteBstWouldAddEndPuncttrue
\mciteSetBstMidEndSepPunct{\mcitedefaultmidpunct}
{\mcitedefaultendpunct}{\mcitedefaultseppunct}\relax
\EndOfBibitem
\bibitem[Mayer \emph{et~al.}(2007)Mayer, Scully, Hardin, Rowell, and
  McGehee]{LYJMCA5}
A.~C. Mayer, S.~R. Scully, B.~E. Hardin, M.~W. Rowell and M.~D. McGehee,
  \emph{Mater. Today}, 2007, \textbf{10}, 28--33\relax
\mciteBstWouldAddEndPuncttrue
\mciteSetBstMidEndSepPunct{\mcitedefaultmidpunct}
{\mcitedefaultendpunct}{\mcitedefaultseppunct}\relax
\EndOfBibitem
\bibitem[Nelson(2011)]{LYJMCA6}
J.~Nelson, \emph{Mater. Today}, 2011, \textbf{14}, 462--470\relax
\mciteBstWouldAddEndPuncttrue
\mciteSetBstMidEndSepPunct{\mcitedefaultmidpunct}
{\mcitedefaultendpunct}{\mcitedefaultseppunct}\relax
\EndOfBibitem
\bibitem[Junquera and Ghosez(2003)]{ACS3}
J.~Junquera and P.~Ghosez, \emph{Nature}, 2003, \textbf{422}, 506--509\relax
\mciteBstWouldAddEndPuncttrue
\mciteSetBstMidEndSepPunct{\mcitedefaultmidpunct}
{\mcitedefaultendpunct}{\mcitedefaultseppunct}\relax
\EndOfBibitem
\bibitem[Xu \emph{et~al.}(2020)Xu, Liu, Chen, Zhang, Ma, Zhang, Sun, and
  Ji]{xu2020two}
Y.~Xu, Y.~Liu, Y.~Chen, Y.~Zhang, C.~Ma, H.~Zhang, S.~Sun and Y.~Ji, \emph{ACS
  Appl Mater Interfaces}, 2020, \textbf{12}, 58349--58359\relax
\mciteBstWouldAddEndPuncttrue
\mciteSetBstMidEndSepPunct{\mcitedefaultmidpunct}
{\mcitedefaultendpunct}{\mcitedefaultseppunct}\relax
\EndOfBibitem
\bibitem[Wang \emph{et~al.}(2018)Wang, Li, Liang, Nie, and Wang]{2018kagse}
Q.~Wang, J.~Li, Y.~Liang, Y.~Nie and B.~Wang, \emph{ACS Appl. Mater.
  Interfaces}, 2018, \textbf{10}, 41670--41677\relax
\mciteBstWouldAddEndPuncttrue
\mciteSetBstMidEndSepPunct{\mcitedefaultmidpunct}
{\mcitedefaultendpunct}{\mcitedefaultseppunct}\relax
\EndOfBibitem
\bibitem[Novoselov \emph{et~al.}(2004)Novoselov, Geim, Morozov, Jiang, Zhang,
  Dubonos, Grigorieva, and Firsov]{gra2004}
K.~S. Novoselov, A.~K. Geim, S.~V. Morozov, D.-e. Jiang, Y.~Zhang, S.~V.
  Dubonos, I.~V. Grigorieva and A.~A. Firsov, \emph{science}, 2004,
  \textbf{306}, 666--669\relax
\mciteBstWouldAddEndPuncttrue
\mciteSetBstMidEndSepPunct{\mcitedefaultmidpunct}
{\mcitedefaultendpunct}{\mcitedefaultseppunct}\relax
\EndOfBibitem
\bibitem[Li \emph{et~al.}(2016)Li, Lin, Lin, Xu, Wang, Zhang, Zhong, Xu, Wu,
  and Fang]{2016h-BN}
X.~Li, S.~Lin, X.~Lin, Z.~Xu, P.~Wang, S.~Zhang, H.~Zhong, W.~Xu, Z.~Wu and
  W.~Fang, \emph{Opt Express}, 2016, \textbf{24}, 134--145\relax
\mciteBstWouldAddEndPuncttrue
\mciteSetBstMidEndSepPunct{\mcitedefaultmidpunct}
{\mcitedefaultendpunct}{\mcitedefaultseppunct}\relax
\EndOfBibitem
\bibitem[Katoch \emph{et~al.}(2018)Katoch, Ulstrup, Koch, Moser, McCreary,
  Singh, Xu, Jonker, Kawakami, Bostwick,\emph{et~al.}]{2018h-BN}
J.~Katoch, S.~Ulstrup, R.~J. Koch, S.~Moser, K.~M. McCreary, S.~Singh, J.~Xu,
  B.~T. Jonker, R.~K. Kawakami, A.~Bostwick \emph{et~al.}, \emph{Nature
  Physics}, 2018, \textbf{14}, 355--359\relax
\mciteBstWouldAddEndPuncttrue
\mciteSetBstMidEndSepPunct{\mcitedefaultmidpunct}
{\mcitedefaultendpunct}{\mcitedefaultseppunct}\relax
\EndOfBibitem
\bibitem[Gogotsi and Anasori(2019)]{2019MXenes}
Y.~Gogotsi and B.~Anasori, \emph{The rise of MXenes}, 2019\relax
\mciteBstWouldAddEndPuncttrue
\mciteSetBstMidEndSepPunct{\mcitedefaultmidpunct}
{\mcitedefaultendpunct}{\mcitedefaultseppunct}\relax
\EndOfBibitem
\bibitem[Chai \emph{et~al.}(2017)Chai, Yan, Wang, Ren, and Zhu]{2017TMDS}
B.~Chai, J.~Yan, C.~Wang, Z.~Ren and Y.~Zhu, \emph{Appl Surf Sci}, 2017,
  \textbf{391}, 376--383\relax
\mciteBstWouldAddEndPuncttrue
\mciteSetBstMidEndSepPunct{\mcitedefaultmidpunct}
{\mcitedefaultendpunct}{\mcitedefaultseppunct}\relax
\EndOfBibitem
\bibitem[Li \emph{et~al.}(2018)Li, Cao, Wang, Xiao, Li, Delaunay, and
  Zhu]{2018TMDS}
C.~Li, Q.~Cao, F.~Wang, Y.~Xiao, Y.~Li, J.-J. Delaunay and H.~Zhu, \emph{Chem
  Soc Rev}, 2018, \textbf{47}, 4981--5037\relax
\mciteBstWouldAddEndPuncttrue
\mciteSetBstMidEndSepPunct{\mcitedefaultmidpunct}
{\mcitedefaultendpunct}{\mcitedefaultseppunct}\relax
\EndOfBibitem
\bibitem[Mak \emph{et~al.}(2010)Mak, Lee, Hone, Shan, and Heinz]{mak2010}
K.~F. Mak, C.~Lee, J.~Hone, J.~Shan and T.~F. Heinz, \emph{Phys Rev Lett},
  2010, \textbf{105}, 136805\relax
\mciteBstWouldAddEndPuncttrue
\mciteSetBstMidEndSepPunct{\mcitedefaultmidpunct}
{\mcitedefaultendpunct}{\mcitedefaultseppunct}\relax
\EndOfBibitem
\bibitem[Novoselov \emph{et~al.}(2016)Novoselov, Mishchenko, Carvalho, and
  Castro~Neto]{20162d}
K.~S. Novoselov, A.~Mishchenko, o.~A. Carvalho and A.~Castro~Neto,
  \emph{Science}, 2016, \textbf{353}, aac9439\relax
\mciteBstWouldAddEndPuncttrue
\mciteSetBstMidEndSepPunct{\mcitedefaultmidpunct}
{\mcitedefaultendpunct}{\mcitedefaultseppunct}\relax
\EndOfBibitem
\bibitem[Gao \emph{et~al.}(2020)Gao, Liu, Yang, Yang, Wang, Zhang, Liu, Jia,
  Xu, and Ma]{gao20202d}
R.~Gao, H.~Liu, J.~Yang, F.~Yang, T.~Wang, Z.~Zhang, X.~Liu, H.~Jia, B.~Xu and
  H.~Ma, \emph{Appl Surf Sci}, 2020, \textbf{529}, 147026\relax
\mciteBstWouldAddEndPuncttrue
\mciteSetBstMidEndSepPunct{\mcitedefaultmidpunct}
{\mcitedefaultendpunct}{\mcitedefaultseppunct}\relax
\EndOfBibitem
\bibitem[Zhang \emph{et~al.}(2020)Zhang, Zhang, Wei, Guo, Fan, Ni, Weng, Zha,
  Liu, Tian,\emph{et~al.}]{zhang2020}
R.~Zhang, Y.~Zhang, X.~Wei, T.~Guo, J.~Fan, L.~Ni, Y.~Weng, Z.~Zha, J.~Liu,
  Y.~Tian \emph{et~al.}, \emph{Appl Surf Sci}, 2020, \textbf{528}, 146782\relax
\mciteBstWouldAddEndPuncttrue
\mciteSetBstMidEndSepPunct{\mcitedefaultmidpunct}
{\mcitedefaultendpunct}{\mcitedefaultseppunct}\relax
\EndOfBibitem
\bibitem[Mahmoud \emph{et~al.}(2019)Mahmoud, Rugut, Molepo, and
  Joubert]{mahmoud2019}
M.~M. Mahmoud, E.~K. Rugut, M.~P. Molepo and D.~P. Joubert, \emph{EUR PHYS J
  B}, 2019, \textbf{92}, 1--10\relax
\mciteBstWouldAddEndPuncttrue
\mciteSetBstMidEndSepPunct{\mcitedefaultmidpunct}
{\mcitedefaultendpunct}{\mcitedefaultseppunct}\relax
\EndOfBibitem
\bibitem[Savelsberg and Sch{\"a}fer(1981)]{1981beitrage}
G.~Savelsberg and H.~Sch{\"a}fer, \emph{Journal of the Less Common Metals},
  1981, \textbf{80}, P59--P69\relax
\mciteBstWouldAddEndPuncttrue
\mciteSetBstMidEndSepPunct{\mcitedefaultmidpunct}
{\mcitedefaultendpunct}{\mcitedefaultseppunct}\relax
\EndOfBibitem
\bibitem[Xu \emph{et~al.}(2019)Xu, Wang, Zheng, Wu, and Xu]{xu2019new}
W.~Xu, R.~Wang, B.~Zheng, X.~Wu and H.~Xu, \emph{ACS Appl. Mater. Interfaces.},
  2019, \textbf{11}, 14457--14462\relax
\mciteBstWouldAddEndPuncttrue
\mciteSetBstMidEndSepPunct{\mcitedefaultmidpunct}
{\mcitedefaultendpunct}{\mcitedefaultseppunct}\relax
\EndOfBibitem
\bibitem[Gobbi \emph{et~al.}(2018)Gobbi, Orgiu, and Samor{\`\i}]{gobbi20182d}
M.~Gobbi, E.~Orgiu and P.~Samor{\`\i}, \emph{Adv Mater}, 2018, \textbf{30},
  1706103\relax
\mciteBstWouldAddEndPuncttrue
\mciteSetBstMidEndSepPunct{\mcitedefaultmidpunct}
{\mcitedefaultendpunct}{\mcitedefaultseppunct}\relax
\EndOfBibitem
\bibitem[Wang \emph{et~al.}(2019)Wang, Li, Liang, Wang, and Nie]{WQJMCA}
Q.~Wang, J.~Li, Y.~Liang, B.~Wang and Y.~Nie, \emph{J. Mater. Chem. A}, 2019,
  \textbf{7}, 10684--10695\relax
\mciteBstWouldAddEndPuncttrue
\mciteSetBstMidEndSepPunct{\mcitedefaultmidpunct}
{\mcitedefaultendpunct}{\mcitedefaultseppunct}\relax
\EndOfBibitem
\bibitem[Su \emph{et~al.}(2020)Su, Li, Hu, Song, Liu, Guo, Zhu, Liu, and
  Pan]{su2020}
Q.~Su, Y.~Li, R.~Hu, F.~Song, S.~Liu, C.~Guo, S.~Zhu, W.~Liu and J.~Pan,
  \emph{Adv Sustain Syst}, 2020, \textbf{4}, 2000130\relax
\mciteBstWouldAddEndPuncttrue
\mciteSetBstMidEndSepPunct{\mcitedefaultmidpunct}
{\mcitedefaultendpunct}{\mcitedefaultseppunct}\relax
\EndOfBibitem
\bibitem[Yang \emph{et~al.}(2018)Yang, Li, Yu, Huang, Ma, and Dai]{yang2018}
H.~Yang, J.~Li, L.~Yu, B.~Huang, Y.~Ma and Y.~Dai, \emph{J. Mater. Chem. A},
  2018, \textbf{6}, 4161--4166\relax
\mciteBstWouldAddEndPuncttrue
\mciteSetBstMidEndSepPunct{\mcitedefaultmidpunct}
{\mcitedefaultendpunct}{\mcitedefaultseppunct}\relax
\EndOfBibitem
\bibitem[Xu \emph{et~al.}(2020)Xu, Zhang, Cheng, Fan, and Yu]{xu2020s}
Q.~Xu, L.~Zhang, B.~Cheng, J.~Fan and J.~Yu, \emph{Chem}, 2020, \textbf{6},
  1543--1559\relax
\mciteBstWouldAddEndPuncttrue
\mciteSetBstMidEndSepPunct{\mcitedefaultmidpunct}
{\mcitedefaultendpunct}{\mcitedefaultseppunct}\relax
\EndOfBibitem
\bibitem[Wang \emph{et~al.}(2022)Wang, Liang, Yao, Li, Liu, Frauenheim, Wang,
  and Wang]{WQFEPV}
Q.~Wang, Y.~Liang, H.~Yao, J.~Li, T.~Liu, T.~Frauenheim, B.~Wang and J.~Wang,
  \emph{J.Mater.Chem.C}, 2022, \textbf{10}, 1048--1061\relax
\mciteBstWouldAddEndPuncttrue
\mciteSetBstMidEndSepPunct{\mcitedefaultmidpunct}
{\mcitedefaultendpunct}{\mcitedefaultseppunct}\relax
\EndOfBibitem
\bibitem[Zhou \emph{et~al.}(2016)Zhou, Zhang, Zhuo, Kou, Ma, Shao, Du, Meng,
  and Frauenheim]{zhou2016}
L.~Zhou, J.~Zhang, Z.~Zhuo, L.~Kou, W.~Ma, B.~Shao, A.~Du, S.~Meng and
  T.~Frauenheim, \emph{J Phys Chem Lett}, 2016, \textbf{7}, 1880--1887\relax
\mciteBstWouldAddEndPuncttrue
\mciteSetBstMidEndSepPunct{\mcitedefaultmidpunct}
{\mcitedefaultendpunct}{\mcitedefaultseppunct}\relax
\EndOfBibitem
\bibitem[Gajdo{\v{s}} \emph{et~al.}(2006)Gajdo{\v{s}}, Hummer, Kresse,
  Furthm{\"u}ller, and Bechstedt]{gajdovs2006}
M.~Gajdo{\v{s}}, K.~Hummer, G.~Kresse, J.~Furthm{\"u}ller and F.~Bechstedt,
  \emph{Phys. Rev. B}, 2006, \textbf{73}, 045112\relax
\mciteBstWouldAddEndPuncttrue
\mciteSetBstMidEndSepPunct{\mcitedefaultmidpunct}
{\mcitedefaultendpunct}{\mcitedefaultseppunct}\relax
\EndOfBibitem
\bibitem[Yang \emph{et~al.}(2016)Yang, Wang, Ataca, Li, Chen, Cai, Suslu,
  Grossman, Jiang, Liu,\emph{et~al.}]{yang2016}
S.~Yang, C.~Wang, C.~Ataca, Y.~Li, H.~Chen, H.~Cai, A.~Suslu, J.~C. Grossman,
  C.~Jiang, Q.~Liu \emph{et~al.}, \emph{ACS Appl Mater Interfaces}, 2016,
  \textbf{8}, 2533--2539\relax
\mciteBstWouldAddEndPuncttrue
\mciteSetBstMidEndSepPunct{\mcitedefaultmidpunct}
{\mcitedefaultendpunct}{\mcitedefaultseppunct}\relax
\EndOfBibitem
\bibitem[Wang \emph{et~al.}(2020)Wang, Liang, Yao, Li, Wang, and Wang]{WQJMCC}
Q.~Wang, Y.~Liang, H.~Yao, J.~Li, B.~Wang and J.~Wang, \emph{J. Mater. Chem.
  C}, 2020, \textbf{8}, 8107--8119\relax
\mciteBstWouldAddEndPuncttrue
\mciteSetBstMidEndSepPunct{\mcitedefaultmidpunct}
{\mcitedefaultendpunct}{\mcitedefaultseppunct}\relax
\EndOfBibitem
\bibitem[Li \emph{et~al.}(2018)Li, Jia, Du, Song, Xia, Wei, and Li]{18type}
X.~Li, G.~Jia, J.~Du, X.~Song, C.~Xia, Z.~Wei and J.~Li, \emph{J Mater Chem C},
  2018, \textbf{6}, 10010--10019\relax
\mciteBstWouldAddEndPuncttrue
\mciteSetBstMidEndSepPunct{\mcitedefaultmidpunct}
{\mcitedefaultendpunct}{\mcitedefaultseppunct}\relax
\EndOfBibitem
\bibitem[Yu \emph{et~al.}(2013)Yu, Liu, Zhou, Yin, Li, Huang, and
  Duan]{yu2013highly}
W.~J. Yu, Y.~Liu, H.~Zhou, A.~Yin, Z.~Li, Y.~Huang and X.~Duan, \emph{Nat
  Nanotechnol}, 2013, \textbf{8}, 952--958\relax
\mciteBstWouldAddEndPuncttrue
\mciteSetBstMidEndSepPunct{\mcitedefaultmidpunct}
{\mcitedefaultendpunct}{\mcitedefaultseppunct}\relax
\EndOfBibitem
\bibitem[Xie \emph{et~al.}(2016)Xie, Zhang, Cai, Huang, Zou, Guo, Gu, and
  Zeng]{xie2016promising}
M.~Xie, S.~Zhang, B.~Cai, Y.~Huang, Y.~Zou, B.~Guo, Y.~Gu and H.~Zeng,
  \emph{Nano Energy}, 2016, \textbf{28}, 433--439\relax
\mciteBstWouldAddEndPuncttrue
\mciteSetBstMidEndSepPunct{\mcitedefaultmidpunct}
{\mcitedefaultendpunct}{\mcitedefaultseppunct}\relax
\EndOfBibitem
\bibitem[Kresse and Hafner(1993)]{JMCA40}
G.~Kresse and J.~Hafner, \emph{Phys Rev B Condens Matter Mater Phys}, 1993,
  \textbf{47}, 558\relax
\mciteBstWouldAddEndPuncttrue
\mciteSetBstMidEndSepPunct{\mcitedefaultmidpunct}
{\mcitedefaultendpunct}{\mcitedefaultseppunct}\relax
\EndOfBibitem
\bibitem[Perdew \emph{et~al.}(1996)Perdew, Burke, and Ernzerhof]{JMCA41}
J.~P. Perdew, K.~Burke and M.~Ernzerhof, \emph{Phys. Rev. Lett}, 1996,
  \textbf{77}, 3865\relax
\mciteBstWouldAddEndPuncttrue
\mciteSetBstMidEndSepPunct{\mcitedefaultmidpunct}
{\mcitedefaultendpunct}{\mcitedefaultseppunct}\relax
\EndOfBibitem
\bibitem[Kresse and Joubert(1999)]{PAW}
G.~Kresse and D.~Joubert, \emph{Phys. Rev. B}, 1999, \textbf{59}, 1758\relax
\mciteBstWouldAddEndPuncttrue
\mciteSetBstMidEndSepPunct{\mcitedefaultmidpunct}
{\mcitedefaultendpunct}{\mcitedefaultseppunct}\relax
\EndOfBibitem
\bibitem[Heyd \emph{et~al.}(2003)Heyd, Scuseria, and Ernzerhof]{JMCA44}
J.~Heyd, G.~E. Scuseria and M.~Ernzerhof, \emph{J. Chem. Phys}, 2003,
  \textbf{118}, 8207--8215\relax
\mciteBstWouldAddEndPuncttrue
\mciteSetBstMidEndSepPunct{\mcitedefaultmidpunct}
{\mcitedefaultendpunct}{\mcitedefaultseppunct}\relax
\EndOfBibitem
\bibitem[Taylor \emph{et~al.}(2001)Taylor, Guo, and Wang]{nanodcal1}
J.~Taylor, H.~Guo and J.~Wang, \emph{Phys. Rev. B}, 2001, \textbf{63},
  245407\relax
\mciteBstWouldAddEndPuncttrue
\mciteSetBstMidEndSepPunct{\mcitedefaultmidpunct}
{\mcitedefaultendpunct}{\mcitedefaultseppunct}\relax
\EndOfBibitem
\bibitem[Waldron \emph{et~al.}(2006)Waldron, Haney, Larade, MacDonald, and
  Guo]{nanodcal2}
D.~Waldron, P.~Haney, B.~Larade, A.~MacDonald and H.~Guo, \emph{Phys. Rev.
  Lett}, 2006, \textbf{96}, 166804\relax
\mciteBstWouldAddEndPuncttrue
\mciteSetBstMidEndSepPunct{\mcitedefaultmidpunct}
{\mcitedefaultendpunct}{\mcitedefaultseppunct}\relax
\EndOfBibitem
\bibitem[Wang \emph{et~al.}(2009)Wang, Wang, and Guo]{KAS122}
B.~Wang, J.~Wang and H.~Guo, \emph{Phys. Rev. B}, 2009, \textbf{79},
  165417\relax
\mciteBstWouldAddEndPuncttrue
\mciteSetBstMidEndSepPunct{\mcitedefaultmidpunct}
{\mcitedefaultendpunct}{\mcitedefaultseppunct}\relax
\EndOfBibitem
\bibitem[Hamann \emph{et~al.}(1979)Hamann, Schl{\"u}ter, and Chiang]{KAS51}
D.~Hamann, M.~Schl{\"u}ter and C.~Chiang, \emph{Phys. Rev. Lett}, 1979,
  \textbf{43}, 1494\relax
\mciteBstWouldAddEndPuncttrue
\mciteSetBstMidEndSepPunct{\mcitedefaultmidpunct}
{\mcitedefaultendpunct}{\mcitedefaultseppunct}\relax
\EndOfBibitem
\bibitem[Perdew \emph{et~al.}(1996)Perdew, Burke, and Ernzerhof]{KAS52}
J.~P. Perdew, K.~Burke and M.~Ernzerhof, \emph{Phys. Rev. Lett}, 1996,
  \textbf{77}, 3865\relax
\mciteBstWouldAddEndPuncttrue
\mciteSetBstMidEndSepPunct{\mcitedefaultmidpunct}
{\mcitedefaultendpunct}{\mcitedefaultseppunct}\relax
\EndOfBibitem
\bibitem[Rold{\'a}n \emph{et~al.}(2015)Rold{\'a}n, Castellanos-Gomez,
  Cappelluti, and Guinea]{2015strain}
R.~Rold{\'a}n, A.~Castellanos-Gomez, E.~Cappelluti and F.~Guinea, \emph{Journal
  of Physics: Condensed Matter}, 2015, \textbf{27}, 313201\relax
\mciteBstWouldAddEndPuncttrue
\mciteSetBstMidEndSepPunct{\mcitedefaultmidpunct}
{\mcitedefaultendpunct}{\mcitedefaultseppunct}\relax
\EndOfBibitem
\bibitem[Lei \emph{et~al.}(2019)Lei, Ma, Xu, Zhang, Huang, and Dai]{jpcc121}
C.~Lei, Y.~Ma, X.~Xu, T.~Zhang, B.~Huang and Y.~Dai, \emph{J. Phys. Chem. C},
  2019, \textbf{123}, 23089--23095\relax
\mciteBstWouldAddEndPuncttrue
\mciteSetBstMidEndSepPunct{\mcitedefaultmidpunct}
{\mcitedefaultendpunct}{\mcitedefaultseppunct}\relax
\EndOfBibitem
\bibitem[Xia \emph{et~al.}(2018)Xia, Du, Li, Li, Zhao, Wang, and Li]{PRA}
C.~Xia, J.~Du, M.~Li, X.~Li, X.~Zhao, T.~Wang and J.~Li, \emph{Phys. Rev.
  Appl}, 2018, \textbf{10}, 054064\relax
\mciteBstWouldAddEndPuncttrue
\mciteSetBstMidEndSepPunct{\mcitedefaultmidpunct}
{\mcitedefaultendpunct}{\mcitedefaultseppunct}\relax
\EndOfBibitem
\bibitem[Ma \emph{et~al.}(2017)Ma, Kuc, Jing, Philipsen, and Heine]{kagse50}
Y.~Ma, A.~Kuc, Y.~Jing, P.~Philipsen and T.~Heine, \emph{Angew Chem Int Ed
  Engl}, 2017, \textbf{56}, 10214--10218\relax
\mciteBstWouldAddEndPuncttrue
\mciteSetBstMidEndSepPunct{\mcitedefaultmidpunct}
{\mcitedefaultendpunct}{\mcitedefaultseppunct}\relax
\EndOfBibitem
\bibitem[Becke and Edgecombe(1990)]{JMCC_SiC2_55}
A.~D. Becke and K.~E. Edgecombe, \emph{J Chem Phys}, 1990, \textbf{92},
  5397--5403\relax
\mciteBstWouldAddEndPuncttrue
\mciteSetBstMidEndSepPunct{\mcitedefaultmidpunct}
{\mcitedefaultendpunct}{\mcitedefaultseppunct}\relax
\EndOfBibitem
\bibitem[Liang \emph{et~al.}(2018)Liang, Dai, Ma, Ju, Wei, and
  Huang]{liang2018}
Y.~Liang, Y.~Dai, Y.~Ma, L.~Ju, W.~Wei and B.~Huang, \emph{J Mater Chem A
  Mater}, 2018, \textbf{6}, 2073--2080\relax
\mciteBstWouldAddEndPuncttrue
\mciteSetBstMidEndSepPunct{\mcitedefaultmidpunct}
{\mcitedefaultendpunct}{\mcitedefaultseppunct}\relax
\EndOfBibitem
\bibitem[Cheng \emph{et~al.}(2018)Cheng, Guo, Han, Jiang, Zhang, Ahuja, Su, and
  Zhao]{JMCA53}
K.~Cheng, Y.~Guo, N.~Han, X.~Jiang, J.~Zhang, R.~Ahuja, Y.~Su and J.~Zhao,
  \emph{Appl Phys Lett}, 2018, \textbf{112}, 143902\relax
\mciteBstWouldAddEndPuncttrue
\mciteSetBstMidEndSepPunct{\mcitedefaultmidpunct}
{\mcitedefaultendpunct}{\mcitedefaultseppunct}\relax
\EndOfBibitem
\bibitem[Bassani \emph{et~al.}(1976)Bassani, Parravicini, Ballinger, and
  Birman]{JMCA560}
F.~Bassani, G.~P. Parravicini, R.~A. Ballinger and J.~L. Birman, \emph{Phys.
  Today}, 1976, \textbf{29}, 58\relax
\mciteBstWouldAddEndPuncttrue
\mciteSetBstMidEndSepPunct{\mcitedefaultmidpunct}
{\mcitedefaultendpunct}{\mcitedefaultseppunct}\relax
\EndOfBibitem
\bibitem[Ma \emph{et~al.}(2018)Ma, Wu, Wang, and Wang]{lyJMCA}
X.~Ma, X.~Wu, H.~Wang and Y.~Wang, \emph{J Mater Chem A Mater}, 2018,
  \textbf{6}, 2295--2301\relax
\mciteBstWouldAddEndPuncttrue
\mciteSetBstMidEndSepPunct{\mcitedefaultmidpunct}
{\mcitedefaultendpunct}{\mcitedefaultseppunct}\relax
\EndOfBibitem
\bibitem[Jin \emph{et~al.}(2017)Jin, Li, Dai, and Wei]{Jinhaopccp}
H.~Jin, J.~Li, Y.~Dai and Y.~Wei, \emph{Phys. Chem. Chem. Phys}, 2017,
  \textbf{19}, 4855--4860\relax
\mciteBstWouldAddEndPuncttrue
\mciteSetBstMidEndSepPunct{\mcitedefaultmidpunct}
{\mcitedefaultendpunct}{\mcitedefaultseppunct}\relax
\EndOfBibitem
\bibitem[Bardeen and Shockley(1950)]{JMCA631}
J.~Bardeen and W.~Shockley, \emph{Phys Rev E}, 1950, \textbf{80}, 72\relax
\mciteBstWouldAddEndPuncttrue
\mciteSetBstMidEndSepPunct{\mcitedefaultmidpunct}
{\mcitedefaultendpunct}{\mcitedefaultseppunct}\relax
\EndOfBibitem
\bibitem[Scharber \emph{et~al.}(2006)Scharber, M{\"u}hlbacher, Koppe, Denk,
  Waldauf, Heeger, and Brabec]{2006design}
M.~C. Scharber, D.~M{\"u}hlbacher, M.~Koppe, P.~Denk, C.~Waldauf, A.~J. Heeger
  and C.~J. Brabec, \emph{Adv Mater}, 2006, \textbf{18}, 789--794\relax
\mciteBstWouldAddEndPuncttrue
\mciteSetBstMidEndSepPunct{\mcitedefaultmidpunct}
{\mcitedefaultendpunct}{\mcitedefaultseppunct}\relax
\EndOfBibitem
\bibitem[Li \emph{et~al.}(2023)Li, Bao, Guo, Ye, Chen, Hou, and Ma]{li2023type}
X.~Li, A.~Bao, X.~Guo, S.~Ye, M.~Chen, S.~Hou and X.~Ma, \emph{Appl Surf Sci},
  2023, \textbf{618}, 156544\relax
\mciteBstWouldAddEndPuncttrue
\mciteSetBstMidEndSepPunct{\mcitedefaultmidpunct}
{\mcitedefaultendpunct}{\mcitedefaultseppunct}\relax
\EndOfBibitem
\bibitem[Shao \emph{et~al.}(2013)Shao, Ye, Yang, and Wang]{shao2013}
Z.-G. Shao, X.-S. Ye, L.~Yang and C.-L. Wang, \emph{J Appl Phys}, 2013,
  \textbf{114}, 093712\relax
\mciteBstWouldAddEndPuncttrue
\mciteSetBstMidEndSepPunct{\mcitedefaultmidpunct}
{\mcitedefaultendpunct}{\mcitedefaultseppunct}\relax
\EndOfBibitem
\bibitem[Radisavljevic \emph{et~al.}(2011)Radisavljevic, Radenovic, Brivio,
  Giacometti, and Kis]{2011single}
B.~Radisavljevic, A.~Radenovic, J.~Brivio, V.~Giacometti and A.~Kis, \emph{Nat
  Nanotechnol}, 2011, \textbf{6}, 147--150\relax
\mciteBstWouldAddEndPuncttrue
\mciteSetBstMidEndSepPunct{\mcitedefaultmidpunct}
{\mcitedefaultendpunct}{\mcitedefaultseppunct}\relax
\EndOfBibitem
\bibitem[Jin \emph{et~al.}(2016)Jin, Li, Wang, Yu, Wan, Xu, Dai, Wei, and
  Guo]{jin2016}
H.~Jin, J.~Li, B.~Wang, Y.~Yu, L.~Wan, F.~Xu, Y.~Dai, Y.~Wei and H.~Guo,
  \emph{J Mater Chem C Mater}, 2016, \textbf{4}, 11253--11260\relax
\mciteBstWouldAddEndPuncttrue
\mciteSetBstMidEndSepPunct{\mcitedefaultmidpunct}
{\mcitedefaultendpunct}{\mcitedefaultseppunct}\relax
\EndOfBibitem
\bibitem[Xia \emph{et~al.}(2018)Xia, Du, Huang, Xiao, Xiong, Wang, Wei, Jia,
  Shi, and Li]{InSePRB}
C.-x. Xia, J.~Du, X.-w. Huang, W.-b. Xiao, W.-q. Xiong, T.-x. Wang, Z.-m. Wei,
  Y.~Jia, J.-j. Shi and J.-b. Li, \emph{Phys. Rev. B}, 2018, \textbf{97},
  115416\relax
\mciteBstWouldAddEndPuncttrue
\mciteSetBstMidEndSepPunct{\mcitedefaultmidpunct}
{\mcitedefaultendpunct}{\mcitedefaultseppunct}\relax
\EndOfBibitem
\bibitem[Chen \emph{et~al.}(2019)Chen, Lei, Wang, Zhong, Liu, Xu, and
  Ouyang]{chen2019tunable}
D.~Chen, X.~Lei, Y.~Wang, S.~Zhong, G.~Liu, B.~Xu and C.~Ouyang, \emph{Appl
  Surf Sci}, 2019, \textbf{497}, 143809\relax
\mciteBstWouldAddEndPuncttrue
\mciteSetBstMidEndSepPunct{\mcitedefaultmidpunct}
{\mcitedefaultendpunct}{\mcitedefaultseppunct}\relax
\EndOfBibitem
\bibitem[Nakamura \emph{et~al.}(1995)Nakamura, Senoh, Iwasa, and
  Nagahama]{1995high}
S.~Nakamura, M.~Senoh, N.~Iwasa and S.-i. N. S.-i. Nagahama, \emph{Jpn J Appl
  Phys}, 1995, \textbf{34}, L797\relax
\mciteBstWouldAddEndPuncttrue
\mciteSetBstMidEndSepPunct{\mcitedefaultmidpunct}
{\mcitedefaultendpunct}{\mcitedefaultseppunct}\relax
\EndOfBibitem
\bibitem[Arakawa and Yariv(1986)]{1986quantum}
Y.~Arakawa and A.~Yariv, \emph{IEEE J Quantum Electron}, 1986, \textbf{22},
  1887--1899\relax
\mciteBstWouldAddEndPuncttrue
\mciteSetBstMidEndSepPunct{\mcitedefaultmidpunct}
{\mcitedefaultendpunct}{\mcitedefaultseppunct}\relax
\EndOfBibitem
\bibitem[Zhao \emph{et~al.}(2016)Zhao, Li, Yang, Jiang, Lin, Ade, Ma, and
  Yan]{JMCA61}
J.~Zhao, Y.~Li, G.~Yang, K.~Jiang, H.~Lin, H.~Ade, W.~Ma and H.~Yan,
  \emph{Nature Energy}, 2016, \textbf{1}, 1--7\relax
\mciteBstWouldAddEndPuncttrue
\mciteSetBstMidEndSepPunct{\mcitedefaultmidpunct}
{\mcitedefaultendpunct}{\mcitedefaultseppunct}\relax
\EndOfBibitem
\bibitem[Tsai \emph{et~al.}(2014)Tsai, Su, Chang, Tsai, Chen, Wu, Li, Chen, and
  He]{JMCA62}
M.-L. Tsai, S.-H. Su, J.-K. Chang, D.-S. Tsai, C.-H. Chen, C.-I. Wu, L.-J. Li,
  L.-J. Chen and J.-H. He, \emph{ACS Nano}, 2014, \textbf{8}, 8317--8322\relax
\mciteBstWouldAddEndPuncttrue
\mciteSetBstMidEndSepPunct{\mcitedefaultmidpunct}
{\mcitedefaultendpunct}{\mcitedefaultseppunct}\relax
\EndOfBibitem
\bibitem[Zhou \emph{et~al.}(2020)Zhou, Gong, Jiang, Xu, Shang, Zhang, Hu, and
  Chu]{zhoubinJMCC}
B.~Zhou, S.-J. Gong, K.~Jiang, L.~Xu, L.~Shang, J.~Zhang, Z.~Hu and J.~Chu,
  \emph{J Mater. Chem. C}, 2020, \textbf{8}, 89--97\relax
\mciteBstWouldAddEndPuncttrue
\mciteSetBstMidEndSepPunct{\mcitedefaultmidpunct}
{\mcitedefaultendpunct}{\mcitedefaultseppunct}\relax
\EndOfBibitem
\bibitem[Zhang \emph{et~al.}(2014)Zhang, Gong, Chen, Liu, Zhu, Xiao, and
  Guo]{JMCA65}
L.~Zhang, K.~Gong, J.~Chen, L.~Liu, Y.~Zhu, D.~Xiao and H.~Guo, \emph{Phys.
  Rev. B}, 2014, \textbf{90}, 195428\relax
\mciteBstWouldAddEndPuncttrue
\mciteSetBstMidEndSepPunct{\mcitedefaultmidpunct}
{\mcitedefaultendpunct}{\mcitedefaultseppunct}\relax
\EndOfBibitem
\bibitem[Xie \emph{et~al.}(2015)Xie, Zhang, Zhu, Liu, and Guo]{JMCA66}
Y.~Xie, L.~Zhang, Y.~Zhu, L.~Liu and H.~Guo, \emph{Nanotechnology}, 2015,
  \textbf{26}, 455202\relax
\mciteBstWouldAddEndPuncttrue
\mciteSetBstMidEndSepPunct{\mcitedefaultmidpunct}
{\mcitedefaultendpunct}{\mcitedefaultseppunct}\relax
\EndOfBibitem
\bibitem[Henrickson(2002)]{JMCA67}
L.~E. Henrickson, \emph{J. Appl. Phys}, 2002, \textbf{91}, 6273--6281\relax
\mciteBstWouldAddEndPuncttrue
\mciteSetBstMidEndSepPunct{\mcitedefaultmidpunct}
{\mcitedefaultendpunct}{\mcitedefaultseppunct}\relax
\EndOfBibitem
\bibitem[Zhao \emph{et~al.}(2017)Zhao, Yang, Li, Jin, Wei, Yu, Huang, and
  Dai]{zhao2017design}
P.~Zhao, H.~Yang, J.~Li, H.~Jin, W.~Wei, L.~Yu, B.~Huang and Y.~Dai, \emph{J
  Mater Chem A Mater}, 2017, \textbf{5}, 24145--24152\relax
\mciteBstWouldAddEndPuncttrue
\mciteSetBstMidEndSepPunct{\mcitedefaultmidpunct}
{\mcitedefaultendpunct}{\mcitedefaultseppunct}\relax
\EndOfBibitem
\bibitem[Jin \emph{et~al.}(2016)Jin, Li, Wang, Yu, Wan, Xu, Dai, Wei, and
  Guo]{jinHao88}
H.~Jin, J.~Li, B.~Wang, Y.~Yu, L.~Wan, F.~Xu, Y.~Dai, Y.~Wei and H.~Guo,
  \emph{J. Mater. Chem. C}, 2016, \textbf{4}, 11253--11260\relax
\mciteBstWouldAddEndPuncttrue
\mciteSetBstMidEndSepPunct{\mcitedefaultmidpunct}
{\mcitedefaultendpunct}{\mcitedefaultseppunct}\relax
\EndOfBibitem
\end{mcitethebibliography}
\bibliographystyle{rsc} %the RSC's .bst file

\end{document}